\documentclass[review,authoryear]{elsarticle}

\usepackage{epsfig, epsf, graphicx, subfigure}
\usepackage{pstricks, pst-node, psfrag}
\usepackage{amssymb,amsmath,bm}
\usepackage{verbatim,enumerate}
\usepackage{rotating, lscape}
\usepackage{setspace}
\usepackage{enumitem}
\usepackage{booktabs}
\usepackage{lscape}
\usepackage{hhline}
\usepackage[T1]{fontenc}
\usepackage{tabularx,ragged2e,booktabs,caption}
\usepackage{array}
\newcolumntype{L}[1]{>{\raggedright\arraybackslash}p{#1}}
\newcolumntype{C}[1]{>{\centering\arraybackslash}p{#1}}
\newcolumntype{R}[1]{>{\raggedleft\arraybackslash}p{#1}}

\usepackage[english]{babel}
\usepackage{multirow}
\usepackage{theorem}
\usepackage{float}

\newtheorem{thm}{Theorem}
\newtheorem{defn}{Definition}

\journal{Computational Statistics and Data Analysis}

\biboptions{authoryear}

\begin{document}

\begin{frontmatter}

\title{Directional Outlyingness for Multivariate Functional Data}

%\tnotetext[t1]{This research was supported by
%	King Abdullah University of Science and Technology (KAUST).}
%\author{{Wenlin Dai\corref{mycorrespondingauthor}} and Marc G. Genton}
%\address{CEMSE Division,
%	King Abdullah University of Science and Technology,
%	Thuwal 23955-6900, Saudi Arabia.}
%\author[mymainaddress,mysecondaryaddress]{Marc G. Genton}

\author[rvt]{Wenlin Dai\corref{cor1}}
%\ead{wenlin.dai@kaust.edu.sa}
\author[rvt]{Marc G. Genton}
%\ead{marc.genton@kaust.edu.sa}

\cortext[cor1]{Statistics Program,
	King Abdullah University of Science and Technology,
	Thuwal 23955-6900, Saudi Arabia. Email: wenlin.dai@kaust.edu.sa}

\address[rvt]{Statistics Program,
	King Abdullah University of Science and Technology,
	Thuwal 23955-6900, Saudi Arabia.}

\begin{abstract}
The direction of outlyingness is crucial to describing the centrality of multivariate functional data.
Motivated by this idea, classical depth is generalized to directional outlyingness for functional data.
Theoretical properties of functional directional outlyingness are investigated and the total outlyingness can be naturally decomposed into two parts: magnitude outlyingness and shape outlyingness which represent the centrality of a curve for magnitude and shape, respectively.
This decomposition serves as a visualization tool for the centrality of curves.
Furthermore, an outlier detection procedure is proposed based on functional directional outlyingness.
This criterion applies to both univariate and multivariate curves and simulation studies show that it outperforms competing methods.
Weather and electrocardiogram data demonstrate the practical application of our proposed framework.

\end{abstract}


\end{frontmatter}

%\linenumbers

\vskip 10pt
\section{Introduction}
Functional data are frequently observed in various fields of science, including but not limited to meteorology, biology, medicine, and engineering.
Examples of functional data are temperature records from weather stations, curves capturing infant growth, expression levels of genes recorded over time, and hand-writing data, to name but a few.
Over the past two decades, much work has been done that analyzes functional data, among which
\citet{ramsayfunctional} provided various parametric methods, \citet{ferraty2006nonparametric} developed detailed nonparametric techniques, and \citet{horvath2012inference} focused on related inferential methods for functional data. Typically, each observation is a real function, either univariate or multivariate, defined on an interval, $\mathcal{I}$, in $\rm \mathbb{R}$.

Statistical depth is widely utilized in nonparametric inference for functional data, for instance, to estimate the trimmed mean \citep{fraiman2001trimmed}, to classify functional data \citep{lopez2006depth}, to construct functional boxplots \citep{sun2011functional}, and to detect outlying curves \citep{arribas2014shape,hubert2015multivariate}. 
It provides adequate tools for exploratory analysis of functional data in different scientific areas such as human health \citep{mckeague2011analyzing}, medical science \citep{hong2014statistical}, and neural science \citep{ngo2015exploratory}.

The concept of statistical depth was initially proposed to rank multivariate data.
Afterwards, it was generalized to provide an ordering from the center outwards of functional data. 
Existing depth functions for functional data include integrated depth (ID; \citealp{fraiman2001trimmed}), $h$-mode and random projection depth \citep{cuevas2006use}, random Tukey depth and integrated dual depth \citep{cuevas2009depth}, band depth and modified band depth (BD and MBD; \citealp{lopez2009concept}), half-region depth and modified half-region depth (HD and MHD; \citealp{lopez2011half}), spatial depth \citep{chakraborty2014spatial,serfling2017depth}, $L^{\infty}$ depth \citep{long2015depth}, extremal depth \citep{narisetty2016extremal,myllymaki2017global}, and total variation depth \citep{huang2016total} for univariate functional data; weighted modified band depth (WMBD; \citealp{ieva2013depth}), simplicial band depth and modified simplicial band depth (SBD and MSBD; \citealp{lopez2014simplicial}), multivariate functional halfspace depth (MFHD; \citealp{claeskens2014multivariate}), and multivariate functional skew-adjusted projection depth (MFSPD; \citealp{hubert2015multivariate}) for multivariate functional data. \citet{chakraborty2014data} investigated the ``degeneracy'' problem and proved that all curves may have zero depth with probability one in infinite-dimensional function spaces for some existing functional depths (like BD and HD). \citet{nieto2016topologically} proposed a formal definition of statistical depth for functional data from a topological point of view. \citet{gijbels2015consistency}, \citet{nagy2016weak}, and \citet{nagy2016integrated} investigated the consistency of non-integrated and integrated depths for functional data, respectively.

Outlier detection is often a necessary step in preliminary analysis of functional data.
Handling a variety of functional outliers including persistent outliers that are outlying on a large part of the domain, isolated outliers that are outlying for a very short time interval, magnitude outliers and shape outliers, can be challenging; see \citet{hubert2015multivariate}.
Many outlier detection methods for functional data have been developed based on functional depth.
\citet{febrero2008outlier} defined outliers as those curves with depths smaller than a threshold generated through a bootstrap procedure.
\citet{hyndman2010rainbow} proposed several visualization tools and made a comparison of some outlier detection methods for functional data.
\citet{sun2011functional} proposed a now-popular visualization and outlier detection tool named \emph{functional boxplot}; this tool is based on MBD but the band depth can be replaced by other functional depths. \citet{gervini2012outlier} detected outliers using boxplots or histograms of a measure of outlyingness.
\citet{arribas2014shape} proposed an \emph{outliergram} that detects shape outliers by connecting two functional depths.
\citet{hubert2015multivariate} presented a taxonomy of functional outliers and proposed two visualization methods for outlier detection.
Apart from these depth-based methods, other criteria have also been proposed for detecting outliers in functional data \citep{gervini2009detecting,hyndman2010rainbow,yu2012outlier}.

\begin{figure}[t!]
	\begin{center}
		\includegraphics[width=12cm,height=6.3cm]{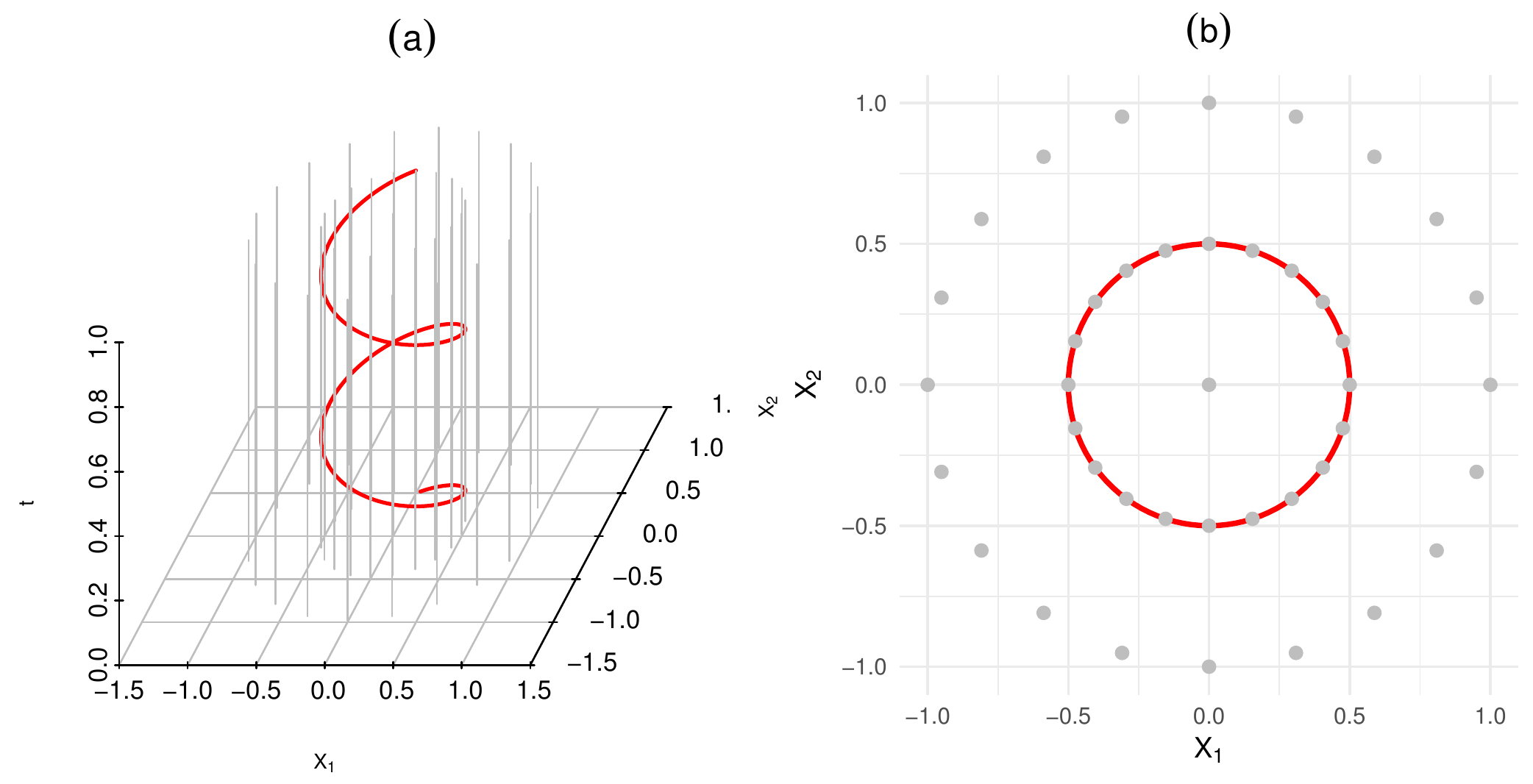}\\
		\caption{(a) Forty-one normal curves (grey) and one outlier (red); (b) Projection of curves along the direction of the $t$-axis. Classical integrated functional depth cannot distinguish between the red and grey curves.}
		\label{motivation}
	\end{center}
\end{figure}

Most existing functional depths, for instance, ID, MBD, MHD, MSBD, MFHD, and MFSPD, fall into the category of integrated functional depth: weighted average of point-wise depth. 
These point-wise depths are always non-negative scalars in the interval $[0,1]$, with large depths assigned to central points and small depths assigned to outward ones.
However, we have found that this type of depths cannot efficiently describe the centrality of functions.
The following example illustrates this deficiency:
42 bivariate curves are defined on $t\in [0,1]$. Among these curves, 41 grey lines take fixed values across $[0,1]$: $(0,0)$, $\{0.5\sin(i\pi/10),0.5\cos(i\pi/10)\}$, and $\{\sin(i\pi/10),\cos(i\pi/10)\}$, $i=1,\dots,20$, respectively, and a red curve is $\{0.5\sin(t\pi/10),0.5\cos(t\pi/10)\}$; see Figure \ref{motivation}(a).
When we project these curves along the $t$ direction, we get Figure \ref{motivation}(b).
Clearly, the red curve is very different from the other curves. It should thus have a smaller depth and be recognized as an outlier.
However, the red curve and the 20 straight lines on the smaller circle actually share the same depth value for some existing functional depths, such as MSBD and MFHD, because they always get the same point-wise depth at each time point $t$.
What makes the red curve an outlier is the change of the direction of outlyingness, which unfortunately has never been considered in existing definitions of functional depths from the literature.

To effectively distinguish the red curve from the grey curves, a natural idea is to consider the direction of outlyingness in addition to the point-wise scalar depth. Bearing this purpose in mind, we propose the framework of directional outlyingness for multivariate functional data, and demonstrate that the new framework enjoys the following three advantages:
\begin{itemize}[noitemsep]
	\item [1.]  Functional directional outlyingness is applicable to both univariate and multivariate functional data with one- or multi- dimensional design domains;
	\item [2.]  Functional directional outlyingness decomposes functional outlyingness into two natural parts: \emph{magnitude outlyingness} and \emph{shape outlyingness}; it provides a much better description of the centrality of curves and the variability among them;
	\item [3.]  An outlier detection criterion is constructed based on functional directional outlyingness; this criterion outperforms existing methods for effectively detecting various types of outliers.
\end{itemize}

The remainder of this paper is organized as follows. In Section 2, we propose the framework of directional outlyingness for multivariate functional data.
In Section 3, we derive an outlyingness decomposition from functional directional outlyingness and visually compare it with classical functional depth. Also, we study important properties and consistency of functional directional outlyingness.
In Section 4, we construct an outlier detection procedure for functional data based on functional directional outlyingness.
In Section 5, we evaluate the proposed outlier detection procedure with Monte Carlo simulation studies for both univariate and multivariate functional data.
We analyze two data examples using our proposed framework in Section 6. We end the paper with a discussion in Section 7. Proofs of the theoretical results are provided in the Appendix.

\section{Functional Directional Outlyingness}

Consider a stochastic process, $\mathbf{X}: \mathcal{I} \longrightarrow \mathbb{R}^p$, that takes values in the space $\mathcal{C}(\mathcal{I},\mathbb{R}^p)$ of real continuous functions defined from a compact interval, $\mathcal{I}$, to $\mathbb{R}^p$ with probability distribution $F_{\mathbf{X}}$. At each fixed time point, $t\in \mathcal{I}$, $\mathbf{X}(t)$ is a $p$-variate random variable with probability distribution $F_{\mathbf{X}(t)}$. Here, $p$ is a finite positive integer that indicates the dimension of the functional data: we get univariate functional data when $p=1$ and multivariate functional data when $p\ge2$. 

Unlike most existing functional depths utilizing statistical depth, our measure of centrality for functional data is built on the concept of \emph{outlyingness}. For multivariate data, outlyingness functions are equivalent to statistical depths in an inverse sense \citep{serfling2006depth}. For functional data, outlyingness turns out to be a better choice, as we demonstrate in the next section. 
Let $d(\mathbf{X}(t),F_{\mathbf{X}(t)})$: $\mathbb{R}^p \longrightarrow [0,1]$ be a statistical depth function for $\mathbf{X}(t)$ with respect to $F_{\mathbf{X}(t)}$. Denote the outlyingness of $\mathbf{X}(t)$ with respect to $F_{\mathbf{X}(t)}$ with $o(\mathbf{X}(t),F_{\mathbf{X}(t)})$. 
The connections among depth, outlyingness, quantile, and rank functions for multivariate data have been comprehensively studied by \citet{zuo2000general} and \citet{serfling2006depth,serfling2010equivariance}. 
Generally, these two quantities can be connected via
\begin{eqnarray}
d(\mathbf{X}(t),F_{\mathbf{X}(t)})=\{1+o(\mathbf{X}(t),F_{\mathbf{X}(t)})\}^{-1}, \label{con1}
\end{eqnarray}
when $o(\mathbf{X}(t),F_{\mathbf{X}(t)})$ has a range $[0,\infty)$, or alternatively via
\begin{eqnarray}
d(\mathbf{X}(t),F_{\mathbf{X}(t)})=1-o(\mathbf{X}(t),F_{\mathbf{X}(t)})/\{\sup o(\cdot,F_{\mathbf{X}(t)})\} \label{con2}
\end{eqnarray}
when $o(\mathbf{X}(t),F_{\mathbf{X}(t)})$ is bounded. \citet{serfling2006depth} introduced three ways to constructing outlyingness for point-type data: projection pursuit approach, distance approach, and quantile function approach. 
Most existing depths can be derived from their corresponding outlyingness; see \citet{zuo2000general} and \citet{serfling2006depth} for discussions. 

To capture the magnitude as well as the direction of outlyingness, we introduce directional outlyingness for multivariate data as follows:
\begin{eqnarray}
\mathbf{O}(\mathbf{X}(t),F_{\mathbf{X}(t)})=o(\mathbf{X}(t),F_{\mathbf{X}(t)})\cdot \mathbf{v}(t)=\left\{{ 1/d(\mathbf{X}(t),F_{\mathbf{X}(t)})}-1\right\}\cdot \mathbf{v}(t), \label{Def1}
\end{eqnarray}
where $d(\mathbf{X}(t),F_{\mathbf{X}(t)})>0$, $\mathbf{v}(t)=\left\{\mathbf{X}(t)-\mathbf{Z}(t)\right\}/\|\mathbf{X}(t)-\mathbf{Z}(t)\|$ is the spatial sign \citep{mottonen1995multivariate} of $\{\mathbf{X}(t)-\mathbf{Z}(t)\}$, $\mathbf{Z}(t)$ denotes the unique median of the distribution $F_{\mathbf{X}(t)}$ with respect to $d(\mathbf{X}(t),F_{\mathbf{X}(t)})$, and $\|\cdot\|$ denotes the $L_2$ norm.  
Here, the magnitude of directional outlyingness is defined based on statistical depth via the connection (\ref{con1}). 
A similar concept in the literature is the centered rank function, $\mathbf{R}(\mathbf{X}(t),F_{\mathbf{X}(t)})$, proposed by \citet{serfling2010equivariance}, which is the inverse of a multivariate quantile function $\mathbf{Q}(\mathbf{X}(t),F_{\mathbf{X}(t)})$. Obviously, the directional outlyingness is different from the centered rank function: $\mathbf{O}(\mathbf{X}(t),F_{\mathbf{X}(t)})$ takes values in $\mathbb{R}^p$ whereas $\mathbf{R}(\mathbf{X}(t),F_{\mathbf{X}(t)})$ takes values in the unit ball $\mathbb{B}^{p-1}(\mathbf{0})\in \mathbb{R}^p$ \citep{serfling2010equivariance}.
For ranking multivariate data, the two quantities are equivalent; however, directional outlyingness is more flexible for ranking functional data. 
Accordingly, the empirical directional outlyingness is defined as $\mathbf{O}_n\left(\mathbf{X}(t),F_{\mathbf{X}(t),n}\right)=\left\{1/d_n(\mathbf{X}(t),F_{\mathbf{X}(t),n})-1\right\}\cdot \mathbf{v}_n(t),$
where $\mathbf{v}_n(t)$ is the unit vector pointing from $\mathbf{Z}_n(t)$ to $\mathbf{X}(t)$ and $\mathbf{Z}_n(t)$ stands for the median of $\{\mathbf{X}_1(t),\dots,\mathbf{X}_n(t)\}$ with respect to the empirical depth $d_n(\mathbf{X}(t),F_{\mathbf{X}(t),n})$.
For the case that $\mathbf{Z}_n(t)$ is not unique, we average all the medians as the unique median for defining $\mathbf{v}_n(t)$. 

Classical depth projects $\mathbf{X}(t)$ onto $[0,1]\subset \mathbb{R}$, whereas directional outlyingness projects $\mathbf{X}(t)$ onto $\mathbb{R}^p$.
We provide an example for the case $p=2$ and calculate the simplicial depth (SD; \citealp{liu1990notion}) and the directional Stahel-Donoho outlyingness (SDO; \citealp{stahel1981breakdown,don0ho_1982breakdown}) for each point to illustrate their difference (Figure \ref{comparison_class_dir}, left column). In comparing classical depth and directional outlyingness, we observe that directional outlyingness retains data positions relative to the median, clearly separating the outliers; whereas, classical depth projects all the outliers onto the left boundary of the plot and consequently mixes them up.
We derive properties of directional outlyingness and present them in the following theorems.
\vskip 5pt
{\thm \label{Theorem 1} For a fixed time point $t_0 \in \mathcal{I}$, assume that $d\left(\mathbf{X}(t_0),F_{\mathbf{X}(t_0)}\right)>0$ is a valid depth function with respect to Definition 2.1 in \citet{zuo2000general}, which means that it possesses four properties: affine invariance, maximality at the center, monotonicity relative to the deepest point and invisibility at infinity. Then, the associated directional outlyingness, $\mathbf{O}\left(\mathbf{X}(t_0),F_{\mathbf{X}(t_0)}\right)$, satisfies the following properties:
	\begin{itemize}[noitemsep]
		\item [\rm (i)] Affine invariance: $\mathbf{O}\left(\mathbf{A}\mathbf{X}(t_0)+\mathbf{b},F_{\mathbf{A}\mathbf{X}(t_0)+\mathbf{b}}\right)=\mathbf{U}\cdot\mathbf{O}\left(\mathbf{X}(t_0),F_{\mathbf{X}(t_0)}\right)$ holds for any $p\times p$ nonsingular matrix, $\mathbf{A}$, and any $p$-vector, $\mathbf{b}$, where $\mathbf{U}=\mathbf{A}\|\mathbf{X}(t_0)-\mathbf{Z}(t_0)\|/\|\mathbf{A}\{\mathbf{X}(t_0)-\mathbf{Z}(t_0)\}\|$. For the norm, we have $\|\mathbf{O}\left(\mathbf{A}\mathbf{X}(t_0)+\mathbf{b},F_{\mathbf{A}\mathbf{X}(t_0)+\mathbf{b}}\right)\|={\|\mathbf{O}\left(\mathbf{X}(t_0),F_{\mathbf{X}(t_0)}\right)\|}$.
		\item [\rm (ii)]Minimality at the center: For any $F_{\mathbf{X}(t_0)}$ with a unique median,  $\mathbf{Z}(t_0)$. $\|\mathbf{O}\left(\mathbf{Z}(t_0),F_{\mathbf{X}(t_0)}\right)\|= \inf_{\mathbf{X}(t_0)\in \mathbb{R}^p}\|\mathbf{O}\left(\mathbf{X}(t_0),F_{\mathbf{X}(t_0)}\right)\|=0$.
		\item [\rm (iii)]Monotonicity relative to the deepest point: For any $F_{\mathbf{X}(t_0)}$ with deepest point $\mathbf{Z}(t_0)$, $\|\mathbf{O}\left(\mathbf{Z}(t_0),F_{\mathbf{X}(t_0)}\right)\|\le \|\mathbf{O}\left(\mathbf{Z}(t_0)+\alpha\{\mathbf{X}(t_0)-\mathbf{Z}(t_0)\},F_{\mathbf{X}(t_0)}\right)\|$ holds for any $\alpha\in [0,1]$.
	\end{itemize}
}
\noindent
When $\textbf{A}$ is an orthogonal matrix, $\|\mathbf{X}(t_0)-\mathbf{Z}(t_0)\|=\|\mathbf{A}\{\mathbf{X}(t_0)-\mathbf{Z}(t_0)\}\|$ and we can rewrite property (i) as $\mathbf{O}\left(\mathbf{A}\mathbf{X}(t_0)+\mathbf{b},F_{\mathbf{A}\mathbf{X}(t_0)+\mathbf{b}}\right)=\mathbf{A}\cdot\mathbf{O}\left(\mathbf{X}(t_0),F_{\mathbf{X}(t_0)}\right)$.

{\thm \label{Consistency of pointwise depth} For a fixed time point $t_0 \in \mathcal{I}$, assume that $$\sup_{\mathbf{X}(t_0)\in \mathbb{R}^p}|d_n\left(\mathbf{X}(t_0),F_{\mathbf{X}(t_0),n}\right)-d\left(\mathbf{X}(t_0),F_{\mathbf{X}(t_0)}\right)|\to 0~a.s.$$ and $\mathbf{Z}_n(t_0)\to \mathbf{Z}(t_0)$ as $n\to \infty$. Then directional outlyingness is consistent:}
$$\sup_{\mathbf{X}(t_0)\in \mathbb{R}^p}\left\|\mathbf{O}_n\left(\mathbf{X}(t_0),F_{\mathbf{X}(t_0),n}\right)-\mathbf{O}\left(\mathbf{X}(t_0),F_{\mathbf{X}(t_0)}\right)\right\|\to 0\quad a.s.\quad {\rm as}~n\to\infty.$$

With the above directional outlyingness for multivariate data, we propose descriptive statistics for functional data. We show that the direction of outlyingness significantly enhances the classical functional depth for illustrating the centrality of functional data.

\noindent
\vskip 5pt
{\defn \label{Definition 2} Consider a stochastic process, $\mathbf{X}: \mathcal{I} \longrightarrow \mathbb{R}^p$, that takes values in the space $\mathcal{C}(\mathcal{I},\mathbb{R}^p)$ of real continuous functions defined from a compact interval, $\mathcal{I}$, to $\mathbb{R}^p$ with probability distribution $F_{\mathbf{X}}$. Let $\mathbf{O}$ be the directional outlyingness defined as (\ref{Def1}) and $w(t)$ a weight function on $\mathcal{I}$. Further assume that, for $t\in \mathcal{I}$, $\mathbf{O}(\mathbf{X}(t),F_{\mathbf{X}(t)}) \in L^2(\mathcal{I}, \mathbb{R}^p)$. We have the following definitions:
	functional directional outlyingness {\rm ({FO})},
	$${\rm FO}(\mathbf{X},F_{\mathbf{X}})=\int_{\mathcal{I}} \|\mathbf{O}(\mathbf{X}(t),F_{\mathbf{X}(t)})\|^2w(t){\rm d}t;$$
	mean directional outlyingness {\rm (\textbf{MO})},
	$$\mathbf{MO}(\mathbf{X},F_{\mathbf{X}})=\int_{\mathcal{I}} \mathbf{O}(\mathbf{X}(t),F_{\mathbf{X}(t)}) w(t){\rm d}t;$$
	and variation of directional outlyingness {\rm (VO)},}
$${\rm VO}(\mathbf{X},F_{\mathbf{X}})=\int_{\mathcal{I}} \|\mathbf{O}(\mathbf{X}(t),F_{\mathbf{X}(t)})-{\rm \mathbf{MO}}(\mathbf{X},F_{\mathbf{X}})\|^2w(t){\rm d}t.$$

\noindent
Here ${\rm FO}$ represents the \emph{total outlyingness} of $\mathbf{X}$, similar to the classical functional depth.
Next, ${\rm \mathbf{MO}}$ describes the relative position (including both distance and direction) of $\mathbf{X}$ on average to the center curve and $\|\mathbf{MO}\|$ can be regarded as the \emph{magnitude outlyingness} of $\mathbf{X}$. Finally, ${\rm VO}$ measures the change of $\mathbf{O}(\mathbf{X}(t),F_{\mathbf{X}(t)})$ in terms of both norm and direction across the whole design interval and can be regarded as the \emph{shape outlyingness} of $\mathbf{X}$.
Unlike classical functional depth, our functional directional outlyingness is no longer a single scalar.
Classical functional depth (fd) is a mapping $\mathbf{X}\in\mathcal{C}(\mathcal{I},\mathbb{R}^p) \longrightarrow {\rm fd}\in[0,1]$, whereas functional directional outlyingness is a mapping $\mathbf{X}\in\mathcal{C}(\mathcal{I},\mathbb{R}^p)\longrightarrow {\rm (\mathbf{MO}^{\rm T},VO)^{\rm T}}\in\mathbb{R}^p\times \mathbb{R}^+$, which greatly enlarges our flexibility of analyzing curves.
The weight function, $w(t)$, can be a constant function \citep{fraiman2001trimmed,lopez2014simplicial}, or proportional to the amount of local variability in amplitude \citep{claeskens2014multivariate}.
Throughout this paper, we use a constant weight function, $w(t)=\{\lambda(\mathcal{I})\}^{-1}$, where $\lambda(\cdot)$ represents the Lebesgue measure.
The finite sample versions of descriptive statistics in Definition \ref{Definition 2} are defined by replacing $F_{\mathbf{X}(t)}$ and $F_{\mathbf{X}}$ with their empirical versions, $F_{\mathbf{X}(t),n}$ and $F_{\mathbf{X},n}$, respectively. For example,
$$\mathbf{MO}_{n}(\mathbf{X},F_{\mathbf{X},n})=\int_{\mathcal{I}} \mathbf{O}_n(\mathbf{X}(t),F_{\mathbf{X}(t),n})\cdot w_n(t){\rm d}t,$$
where $w_n(t)$ is the finite sample version of the chosen weight function and other quantities can be defined accordingly.

Different depth notions can be utilized to define functional directional outlyingness. 
We generally divide point-wise depths into two classes according to their dependence on either rank or distance information. 
		The first class contains rank-based depths and includes half-region depth \citep{tukey1975mathematics} and simplicial depth (SD; \citealp{liu1990notion}) among others; the second class contains distance-based depths and includes Mahalanobis depth \citep{mahalanobis1936generalized}, spatial depth \citep{vardi2000multivariate} and projection depth (PD; \citealp{zuo2003projection}) among others. 
		Through our investigation, we find that a distance-based depth is more appropriate for constructing directional outlyingness because it involves more information than a rank-based depth.
In this paper, we use the projection depth \citep{zuo2003projection} to generate numerical results.
Specifically, the projection depth is defined as
$${\rm PD}(\mathbf{X}(t),F_{\mathbf{X}(t)})=\{1+{\rm SDO}(\mathbf{X}(t),F_{\mathbf{X}(t)})\}^{-1},$$
where
$${\rm SDO}(\mathbf{X}(t),F_{\mathbf{X}(t)})=\sup_{\|\mathbf{u}\|=1}{\|\mathbf{u}^{\rm T}\mathbf{X}(t)-{\rm median}(\mathbf{u}^{\rm T}\mathbf{X}(t))\|\over {\rm MAD}(\mathbf{u}^{\rm T}\mathbf{X}(t))}$$
is the Stahel-Donoho outlyingness \citep{stahel1981breakdown,don0ho_1982breakdown}. Following (\ref{Def1}), the directional SDO $(\mathbf{dSDO})$ is expressed as:
$${\rm \mathbf{dSDO}}(\mathbf{X}(t),F_{\mathbf{X}(t)})={\rm SDO}(\mathbf{X}(t),F_{\mathbf{X}(t)})\cdot\mathbf{v}(t).$$
Substituting \textbf{dSDO} into Definition \ref{Definition 2}, we get a Stahel--Donoho type of functional directional outlyingness $(\mathbf{dFSDO})$.

\section{Properties of Functional Directional Outlyingness}
This section introduces two theorems on the properties of functional directional outlyingness. In particular, we show that the proposed functional directional outlyingness provides a natural decomposition of the functional outlyingness. It also shares some similar properties with classical functional depth.

\subsection{Functional Outlyingness Decomposition}

\begin{figure}[t!]
	\begin{center}
		\includegraphics[width=12cm,height=12cm]{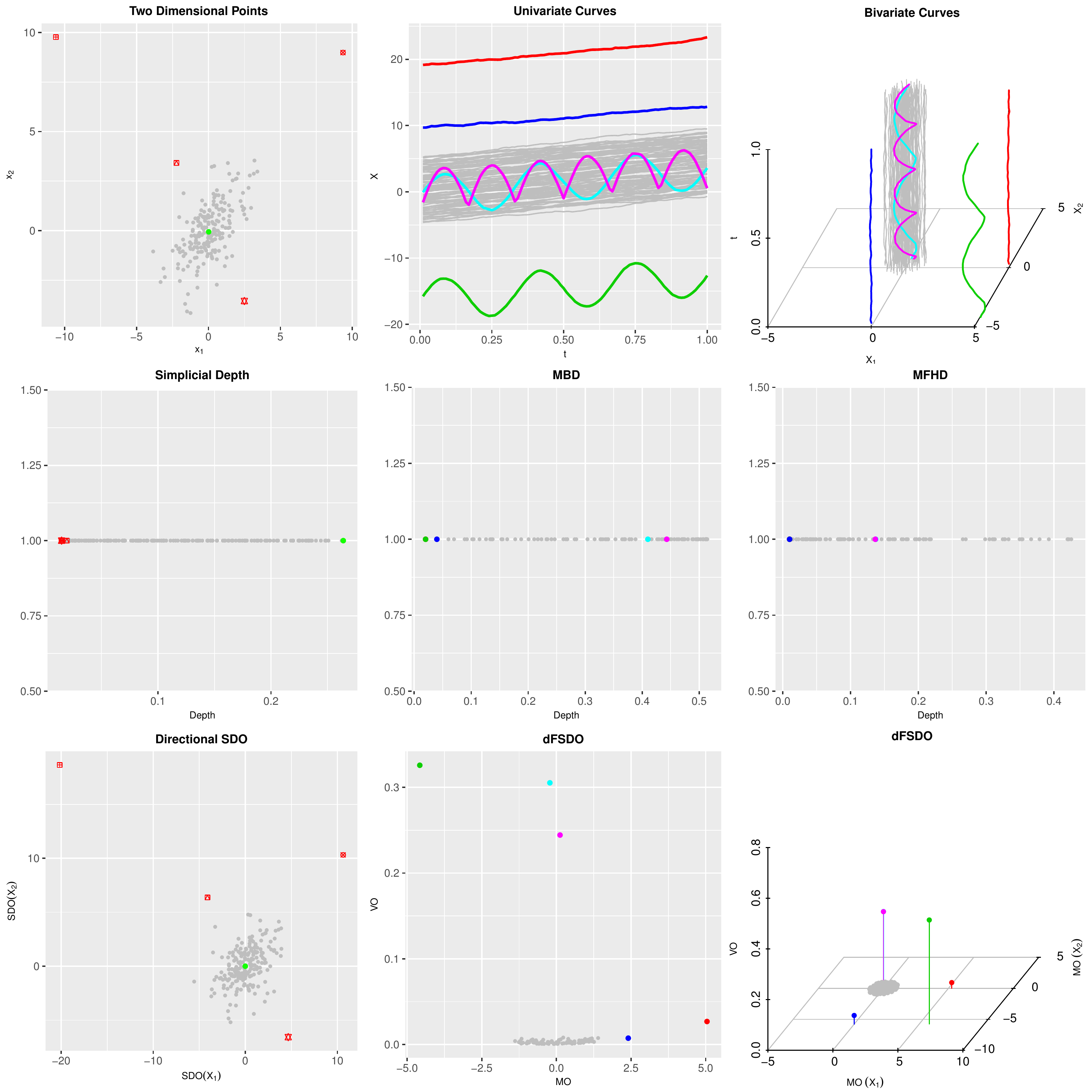}\\
		\caption{Left column: scatter plot of 204 points generated from a normal distribution. The four red points are outliers and the green point is the median with respect to simplical depth (top), simplicial depth (middle) and directional SDO (bottom). Middle column: 100 univariate curves (grey) with 5 outlying curves (various colors), MBD (middle) and directional functional SDO (\textbf{dFSDO}, bottom). Right column: 100 bivariate curves (grey) with 5 outlying curves (various colors), MFHD (middle) and \textbf{dFSDO} (bottom).}
		\label{comparison_class_dir}
	\end{center}
\end{figure}

\vskip 5pt
{\thm \label{Theorem 2} For the proposed statistics in Definition \ref{Definition 2}, we have the following relationship:}
\begin{equation}
{\rm FO}(\mathbf{X},F_{\mathbf{X}})=\|{\rm \mathbf{MO}}(\mathbf{X},F_{\mathbf{X}})\|^2+{\rm VO}(\mathbf{X},F_{\mathbf{X}}). \label{decompose}
\end{equation}
Equation (\ref{decompose}) provides a decomposition of ${\rm FO}$: total functional directional outlyingness is separated into two parts, magnitude outlyingness and shape outlyingness.
When a group of curves shares the same shape or, equivalently, they are parallel with each other, we may expect that ${\rm VO}$ is close to zero. In that case, a quadratic relationship appears between ${\rm FO}$ and $\|\mathbf{MO}\|$.
In contrast, classical depth cannot be decomposed in this way because it does not contain the information of the direction.
To illustrate this difference, we provide examples for both univariate and bivariate curves (Figure \ref{comparison_class_dir}, middle and right columns).

For the univariate case, we generate a sample of curves with different types of outliers and compare the MBD with the \textbf{dFSDO} of each curve (Figure \ref{comparison_class_dir}, middle column). Five outliers are presented in different colors, including two pure magnitude outliers (red and blue), two pure shape outliers (cyan and purple), and one outlier (green) that is outlying for both magnitude and shape.
Two shifted outliers are all mapped to the left edge of the plot and are not easy to be distinguished from the normal curves on the boundary for MBD,
whereas, they are well separated by their much larger MO from \textbf{dFSDO}.
Also, the two shape outliers for MBD are extensively covered with normal curves, whereas they are clearly isolated from the normal curves by their larger VO from \textbf{dFSDO}.
This again indicates that the shape outlyingness (VO) is actually an effective measure of the variation in the shape of a curve.
Moreover, MBD only demonstrates the green curve's outlyingness for magnitude and cannot tell the difference between the green and red curves.
In contrast, \textbf{dFSDO} comprehensively illustrates the green curve's outlyingness for magnitude and shape: a smaller MO stands for its downward shift and a larger VO stands for its different shape.
We compare the MFHD and the \textbf{dFSDO} of the bivariate case (Figure 2, right column) and observe similar improvements.
Overall, with the functional directional outlyingness and the decomposition (\ref{decompose}), we obtain a much better visualization of the curves with respect to the centrality.

%\begin{figure}
%\begin{center}
%\includegraphics[width=16.5cm,height=6cm]{comparison_functionaldepth_ui}\\
%\caption{Left: 100 univariate curves with 5 outlying curves (colored ones) among them. Middle: classical MBD of each curve. Right: directional MBD (\textbf{dMBD}).}
%\label{comparison_functionaldepth_ui}
%\end{center}
%\end{figure}
%
%
%\begin{figure}
%\begin{center}
%\includegraphics[width=16.5cm,height=6cm]{comparison_functionaldepth_bi}\\
%\caption{Left: 100 bivariate curves with 5 outlying curves (colored ones) among them. Middle: classical MFHD of each curve. Right: directional MFHD (\textbf{dMFHD}).}
%\label{comparison_functionaldepth_bi}
%\end{center}
%\end{figure}

\subsection{Theoretical Properties}

Now assume that $\mathcal{I}$ is a compact interval, that the distribution $F_{\mathbf{X}}$ is absolutely continuous and that $F_{\mathbf{X}(t)}$ has a unique median, $\mathbf{Z}(t)$, for each $t\in \mathcal{I}$. We investigate the properties of functional directional outlyingness and present them in the following theorem.
\vskip 5pt
{\thm \label{Theorem 3} Assume that $\mathbf{O}(\mathbf{X}(t),F_{\mathbf{X}(t)})$ is a directional outlyingness from Theorem \ref{Theorem 1}. Then, $\mathbf{MO}$, ${\rm VO}$ and ${\rm FO}$ in Definition \ref{Definition 2} possess the following properties:
	\begin{itemize}[noitemsep]
		\item [1.] Transformation invariance:
		\begin{itemize}[noitemsep]
			\item [(a)] Let $\mathbf{T}(\mathbf{X})$ be a functional, $\mathcal{C}(\mathcal{I},\mathbb{R}^p)\longrightarrow\mathcal{C}(\mathcal{I},\mathbb{R}^p)$, with the form $\mathbf{T}(\mathbf{X}(t))=\mathbf{A}(t)\mathbf{X}(t)+\mathbf{b}(t)$, where $\mathbf{A}(t)$ is a nonsingular matrix and $\mathbf{b}(t)$ is a $p$-vector for each $t\in \mathcal{I}$. In addition, let $g$ be a
			bijection on the interval, $\mathcal{I}$, and let the weight function, $w(t)$, be a constant function. Then,
			\begin{eqnarray*}
				{\rm FO}\left(\mathbf{T}(\mathbf{X}_{g}),F_{\mathbf{T}(\mathbf{X}_{g})}\right)={\rm FO}\left(\mathbf{X},F_{\mathbf{X}}\right),
			\end{eqnarray*}
			where $\mathbf{X}_{g}(t)=\mathbf{X}(g(t))$ for each $t\in \mathcal{I}$.
			\item [(b)] Moreover, if we further assume that $\mathbf{A}(t)=f(t)\mathbf{A}_0$, where $f(t)>0$ for $t\in \mathcal{I}$ and $\mathbf{A}_0$ is an orthogonal matrix, then we have
			\begin{eqnarray*}
				\mathbf{MO}\left(\mathbf{T}(\mathbf{X}_{g}),F_{\mathbf{T}(\mathbf{X}_{g})}\right)&=&\mathbf{A}_0\cdot\mathbf{MO}\left(\mathbf{X},F_{\mathbf{X}}\right),\\
				{\rm VO}\left(\mathbf{T}(\mathbf{X}_{g}),F_{\mathbf{T}(\mathbf{X}_{g})}\right)&=&{\rm VO}\left(\mathbf{X},F_{\mathbf{X}}\right).
			\end{eqnarray*}
		\end{itemize}
		\item [2.] Minimality at the center:
		\begin{eqnarray*}
			{\rm FO}\left(\mathbf{Z},F_{\mathbf{X}}\right)=\inf_{X\in \mathcal{C}(\mathcal{I},\mathbb{R}^p)}{\rm FO}\left(\mathbf{X},F_{\mathbf{X}}\right),
		\end{eqnarray*}
		for any distribution, $F_{\mathbf{X}}$, with a unique center function of symmetry $\mathbf{Z}$.
		\item [3.] Monotonicity with respect to the deepest point: Assume that $\mathbf{Z}$ exists for $F_{\mathbf{X}}$ such that $\mathbf{Z}(t)$ is the deepest point at every $t\in \mathcal{I}$. Then, for any $\alpha\in[0,1]$, we have that
		\begin{eqnarray*}
			{\rm FO}\left(\mathbf{X},F_{\mathbf{X}}\right)\ge {\rm FO}\left(\mathbf{X}+\alpha(\mathbf{Z}-\mathbf{X}),F_{\mathbf{X}}\right).
		\end{eqnarray*}
	\end{itemize}
}
\noindent
If we use an affine invariant weight function proportional to the local variability, we need to further assume that ${\rm det}\{\mathbf{A}(t)\}$ is a constant for each $t\in \mathcal{I}$ to guarantee the affine invariance of ${\rm FO}$. This result is more general than Theorem 1 in \citet{claeskens2014multivariate}, which applies an identical transformation to $\mathbf{X}(t)$ at each time point.

In practice, we get observations only at a finite set of time points, say $T_k=\{t_1,t_2,\dots,t_k\}$ in $\mathcal{I}$, for a finite sample of curves.
Under this setting, we can also define some descriptive statistics as in Definition \ref{Definition 2}.
For example, we may define
\begin{eqnarray*}
	\mathbf{MO}_{T_k}\left(\mathbf{X},F_{\mathbf{X}}\right)&=&{1\over k}\sum_{i=1}^k \mathbf{O}\left(\mathbf{X}(t_i),F_{\mathbf{X}(t_i)}\right)w(t_i),\\
	\mathbf{MO}_{T_k,n}\left(\mathbf{X},F_{\mathbf{X},n}\right)&=&{1\over k}\sum_{i=1}^k \mathbf{O}_n\left(\mathbf{X}(t_i),F_{\mathbf{X}(t_i),n}\right)w_n(t_i).
\end{eqnarray*}
Finite dimensional versions can be defined similarly for the other four statistics.
\vskip 5pt
{\thm \label{Theorem 4}  For the finite-dimensional setting above, assume that the convergence of $\mathbf{O}_n$ holds as stated in Theorem \ref{Consistency of pointwise depth} and that $w_n(t)$ converges almost surely to $w(t)$ as $n\to \infty$ for each $t\in T_k$. Then, we get the convergence of $\mathbf{MO}_{T_k,n}$ as}
$$\sup_{\mathbf{X}\in (\mathbb{R}^p)^k}\|\mathbf{MO}_{T_k,n}\left(\mathbf{X},F_{\mathbf{X},n}\right)-\mathbf{MO}_{T_k}\left(\mathbf{X},F_{\mathbf{X}}\right)\|\to 0\quad a.s.\quad {\rm as}~n\to\infty.$$
When $k=k(n)\to \infty$, one can follow the proof of Theorem 3 in \citet{claeskens2014multivariate} to obtain the consistency of $\mathbf{MO}_{T_k,n}\left(\mathbf{X},F_{\mathbf{X},n}\right)\to \mathbf{MO}\left(\mathbf{X},F_{\mathbf{X}}\right)$.

\section{Functional Outlier Detection}
Apart from better visualizing the centrality of curves, we can also design an outlier detection procedure with functional directional outlyingness.
In the literature, a depth-based outlier detection procedure relies on either a cutoff of functional depth distribution chosen by bootstrap \citep{febrero2008outlier} or an envelop constructed by central curves \citep{hyndman2010rainbow,sun2011functional}.
In the current paper, we design our outlier detection method using $\mathbf{MO}_{T_k,n}$ and ${\rm VO}_{ T_k,n}$ jointly.

%{\thm \label{Theorem 6} Assume that $\mathbf{X}$ is a $p$-dimensional stationary Gaussian process defined on $\mathcal{I}$ and it is $\alpha$-mixing with $\alpha=O(k^{-5})$. 
%Assume the marginal distribution is ${N}_p(\mathbf{0},\mathbf{\Sigma})$ with $\mathbf{\Sigma}>0$ and
%$\mathbf{X}_{i, T_k}$, $i=1,2,\dots$, are idenpendent realizations of this process on the same set of points: ${ T_k}=\{t_1,t_2,\dots,t_k\} \subseteq \mathcal{I}$.
%When the directional outlyingness is constructed with the random projection depth, $\sqrt{k}\mathbf{Y}_{ T_k}=(\sqrt{k}\mathbf{MO}_{T_k}^{\rm T},\sqrt{k}{\rm VO}_{T_k})^{\rm T}$ asymptotically follows a $(p+1)$-dimensional normal distribution as $k\to \infty$.}\\
%The $\alpha$-mixing condition guarantees the asymptotic normality of the vector $\mathbf{Y}_{T_k}=(\mathbf{MO}_{T_k}^{\rm T},{\rm VO}_{T_k})^{\rm T}$. This condition is satisfied by many random processes, for instance, independent, Markov, and autoregressive processes. The proof of Theorem \ref{Theorem 6} is provided in the Appendix. We did not pursue the specific limiting distribution because it is not necessary for implementing the outlier detection method described below. 

%By Theorems 5 and 6, we may approximate the distribution of $\mathbf{Y}_{k,n}=(\mathbf{MO}^{\rm T}_{{ T_k},n},	{\rm VO}_{{T_k},n})^{\rm T}$ with a $(p+1)$-dimensional normal distribution. 
%
%Assume that $\mathbf{X}$ is a $p$-dimensional stationary Gaussian process
%This procedure turns out to be quite effective in our numerical results.
\begin{figure}[b!]
	\begin{center}
		\includegraphics[width=12cm,height=8cm]{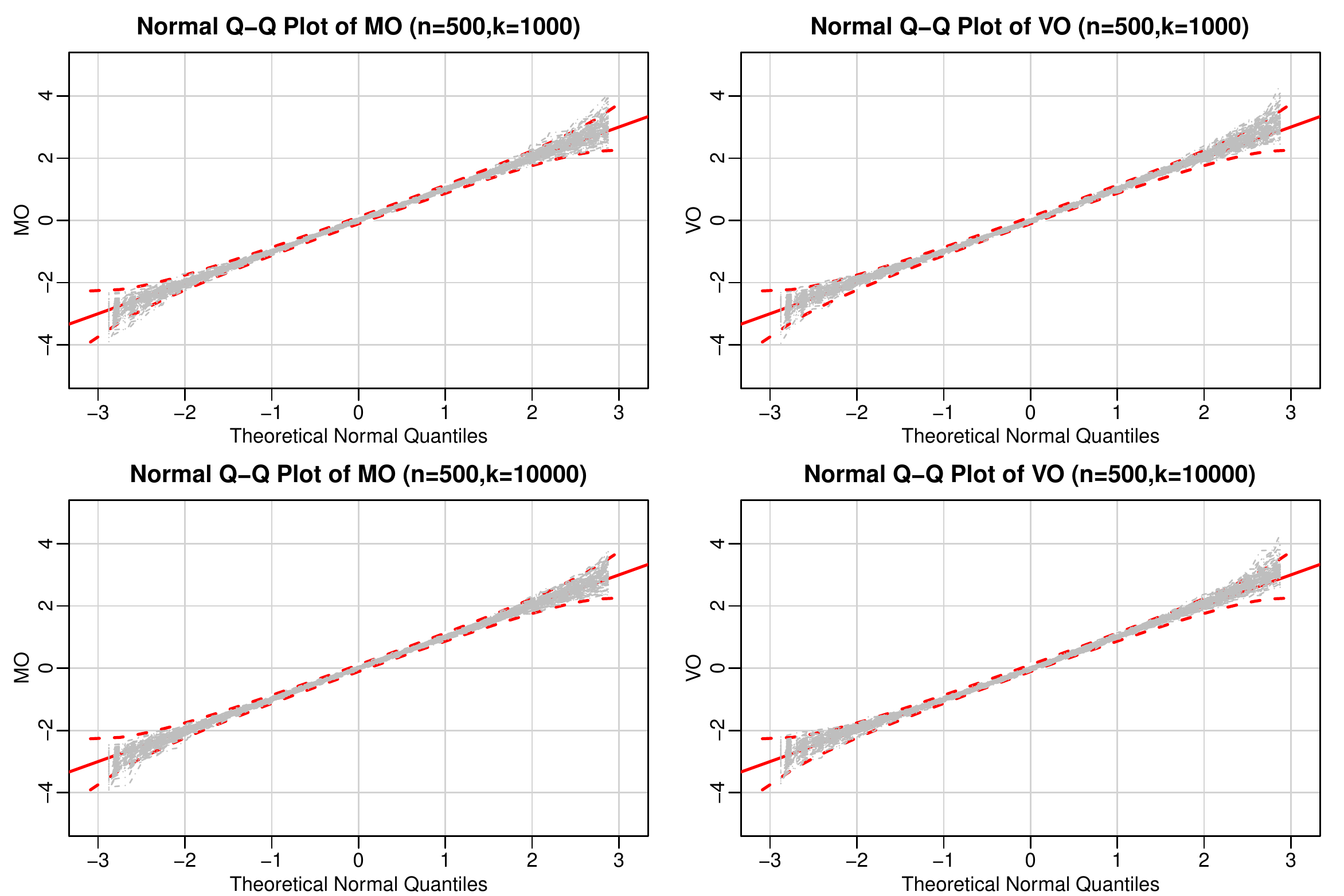}
		\caption{100 normal Q-Q plot trajectories (grey curves) of ${\rm MO}_{T_k,n}$ and ${\rm VO}_{T_k,n}$ under two settings of Model 0 are shown together with a $95\%$ point-wise confidence envelop of the quantiles of the standard normal distribution. The red solid line: $y=x$. The two red dashed lines: upper and lower bounds of the confidence envelop. We have standardized the quantities and the theoretical quantiles are generated from the standard normal distribution. Top panel: $k=1000$; bottom panel: $k=10000$. Both ${\rm MO}_{T_k,n}$ and ${\rm VO}_{T_k,n}$ are well approximated by the normal distribution.}
		\label{qqplot}
	\end{center}
\end{figure}

Through our numerical studies, we found that the distribution of $\mathbf{Y}_{k,n}=(\mathbf{MO}^{\rm T}_{{ T_k},n},	{\rm VO}_{{T_k},n})^{\rm T}$ can be well approximated with a $(p+1)$-dimensional normal distribution when $\mathbf{X}$ is generated from a $p$-dimensional stationary Gaussian process. As an example, we generated 500 curves at $k$ equidistant points on $[0,1]$ from the random process (denoted as Model 0): $X(t)=4t+\epsilon(t)$, where $\epsilon(t)$ is a Gaussian process with zero mean and covariance function $r(s,t)=\exp\{-1000|t-s|\}$. We calculated $\mathbf{MO}_{T_k,n}$ and ${\rm VO}_{T_k,n}$ with different numbers of design points, $k=1000$ and $10000$. Then we computed the normal Q-Q plot trajectory of each quantity. We repeated this procedure 100 times and plot these trajectories under two settings; see Figure \ref{qqplot}. As shown, the two quantities are well approximated by the normal distribution under both settings.

Such an approximation allows us to use the results from \citet{hardin2005distribution} to detect potential outliers from $\mathbf{Y}_{k,n}$. In particular, we calculate the robust Mahalanobis distance of $\mathbf{Y}_{k,n}$ with \cite{rousseeuw1985multivariate}'s minimum covariance determinant estimators for shape and location of the data.
Finally, we employ the approximation proposed by \citet{hardin2005distribution} for the distribution of these distances and determine the cutoff based on such an approximation. 
It turns out that this procedure is quite effective in the simulation studies.
Specifically, the outlier detection is conducted by the following three steps:
\begin{itemize}[noitemsep]
	\item [1.] Calculate the robust Mahalanobis distance based on a sample of size $h\le n$,
	$${\rm RMD}^2(\mathbf{Y}_{k,n},\mathbf{\bar Y^*}_{k,n,{\rm J}})=(\mathbf{Y}_{k,n}-\mathbf{\bar Y^*}_{k,n,{\rm J}})^{\rm T}{\mathbf{S}^*_{{k,n},{\rm J}}}^{-1}(\mathbf{Y}_{k,n}-\mathbf{\bar Y^*}_{{k,n},{\rm J}}),$$
	where $\rm J$ denotes the group of $h$ points that minimizes the determinant of the corresponding covariance matrix, $\mathbf{\bar Y^*}_{{k,n},{\rm J}}=h^{-1}\sum_{i\in \rm J}\mathbf{Y}_{{k,n},i}$ and
	$\mathbf{S}^*_{{k,n},{\rm J}}=h^{-1}\sum_{i\in \rm J}(\mathbf{Y}_{{k,n},{i}}-\mathbf{\bar Y^*}_{{k,n},{\rm J}})(\mathbf{Y}_{{k,n},{i}}-\mathbf{\bar Y^*}_{{k,n},{\rm J}})^{\rm T}$. The sub-sample of size $h$ controls the robustness of the method. For a $p$-dimensional distribution, the maximum finite sample breakdown point is $[(n-p+1)/2]/n$, where $[a]$ denotes the integer part of $a\in \mathbb{R}$.
	\item [2.] According to the results in \citet{hardin2005distribution}, approximate the tail of this distance distribution as follows:
	$${c(m-p)\over m(p+1)}{\rm RMD}^2(\mathbf{Y}_{k,n},\mathbf{\bar Y^*}_{k,n,{\rm J}})\sim F_{p+1,m-p},$$
	where $c$ and $m$ are parameters determining the degrees of freedom of the $F$-distribution and the scaling factor, respectively.
	They are calculated by a simulation program provided by \citet{hardin2005distribution}.
	Then, we choose a cutoff value, $C$, as the $\alpha$ quantile of $F_{p+1,m-p}$. We set $\alpha=0.993$, which is used in the classical boxplot for detecting outliers under a normal distribution.
	\item [3.] Flag a curve as an outlier when its distance satisfies $${c(m-p)\over m(p+1)}{\rm RMD}^2(\mathbf{Y}_{k,n},\mathbf{\bar Y^*}_{k,n,{\rm J}})>C.$$
\end{itemize}

The ${\rm RMD}$ can also serve as a measure of centrality for the curves, based on which we can define the median and the central region of the curves, calculate the trimmed mean function, and generate the functional boxplot as well.

\section{Monte Carlo Simulation Studies}

We have demonstrated the superiority of directional outlyingness in describing and visualizing both point-wise and functional data.
In this section, we conduct simulation studies to assess the performance of the proposed outlier detection procedure based on RMD, denoted as Dir.Out.
Our simulation studies include both univariate and multivariate functional data.
Throughout the simulations, we use a constant weight function for calculating \textbf{dFSDO}.

\subsection{Univariate Functional Data}
For univariate functional data, we compare Dir.Out with three existing methods:
\begin{itemize}[noitemsep]
	\item [1.] Integrated squared error method (Int.Sqe; \citealp{hyndman2007robust}): the integrated squared error is calculated for each curve after extracting some fixed number of principal components. Outliers are defined as those observations with an integrated squared error greater than a threshold. We use \emph{foutliers} in the R package \emph{rainbow} with the default setting.
	%\item [2.] Depth-based trimming \citep{febrero2008outlier}.
	\item [2.] Robust Mahalanobis distance (Rob.Mah; \citealp{hyndman2010rainbow}): the squared robust Mahalanobis distances are calculated by treating the curves as $q$-dimensional data and the resulting distances follow a $\chi^2$ distribution with $q$ degrees of freedom. A curve is flagged as an outlier if its squared distance is larger than $\chi^2_{0.99,q}$, the critical value. We use \emph{foutliers} in the R package \emph{rainbow} with the default setting.
	%\item [4.] Adjusted functional boxplot \citep{sun2012adjusted}.
	\item [3.] Adjusted Outliergram (Out.Grm; \citealp{arribas2014shape}): the functional boxplot is first applied to detect the magnitude outliers and then the shape outliers are detected by the boxplot of a measurement of shape variation. We use the R code \emph{OutGramAdj} from\\ http://halweb.uc3m.es/esp/Personal/personas/aarribas/esp/public.html.
\end{itemize}

We consider four models with different types of outliers and different contamination levels $\epsilon=0$ (uncontaminated), $0.1$ and $0.2$.
\begin{itemize}[noitemsep]
	\item Model 1 (shifted outlier). Main model: $X(t)=4t+e(t)$, where $e(t)$ is a Gaussian process with zero mean and covariance function $\gamma(s,t)=\exp\{-|t-s|\}$. Contamination model: $X(t)=4t+8U+e(t)$, where $U$ takes values $-1$ and $1$ with probability $1/2$.
	\item Model 2 (isolated outlier). Main model: $X(t)=4t+e(t)$ and contamination model: $X(t)=4t+8UI_{\{T\le t\le T+0.1\}}+e(t)$, where $T$ is generated from a uniform distribution on $[0,0.9]$ and $I_{A}$ is an indicator function taking value 1 on set $A$ and 0 otherwise. Models 1 and 2 were considered by \citet{sun2011functional}.
	\item Model 3 (shape outlier I). Main model: $X(t)=30t(1-t)^{3/2}+{\tilde e}(t)$, where ${\tilde{e}}(t)$ is a Gaussian process with zero mean and covariance function $\tilde\gamma(s,t)=0.3\exp\{-|t-s|/0.3\}$. Contamination model: $X(t)=30(1-t)t^{3/2}+\tilde{e}(t)$. This model was considered in \citet{arribas2014shape}.
	\item Model 4 (shape outlier II). Main model: $X(t)=4t+\tilde e_1(t)$, where $\tilde e_1(t)$ is a Gaussian process with zero mean and covariance function $\gamma_1(s,t)=\exp\{-|t-s|\}$. Contamination model: $X(t)=4t+\tilde e_2(t)$, where $\tilde e_2(t)$ is a Gaussian process with zero mean and covariance function $\gamma_2(s,t)=8\exp\{-|t-s|^{0.2}\}$. This model was considered by \citet{sun2011functional}. The contamination is from the covariance function but eventually leads to different shapes for the two groups. 
\end{itemize}
For each case, we generate 100 curves at 50 equidistant time points on $[0,1]$.
For the proposed outlier detection procedure, we set $h=[0.75n]$, which means that we assume that the proportion of outliers in a group of curves is no more than $25\%$.
For comparison, we calculate two quantities: the correct detection rate, $p_c$ (number of correctly detected outliers divided by the total number of outlying curves), and the false detection rate, $p_f$ (number of falsely detected outliers divided by the total number of normal curves).
The simulation results are presented in Table \ref{Table_uni}.

\begin{figure}[b!]
	\begin{center}
		\includegraphics[width=12cm,height=4cm]{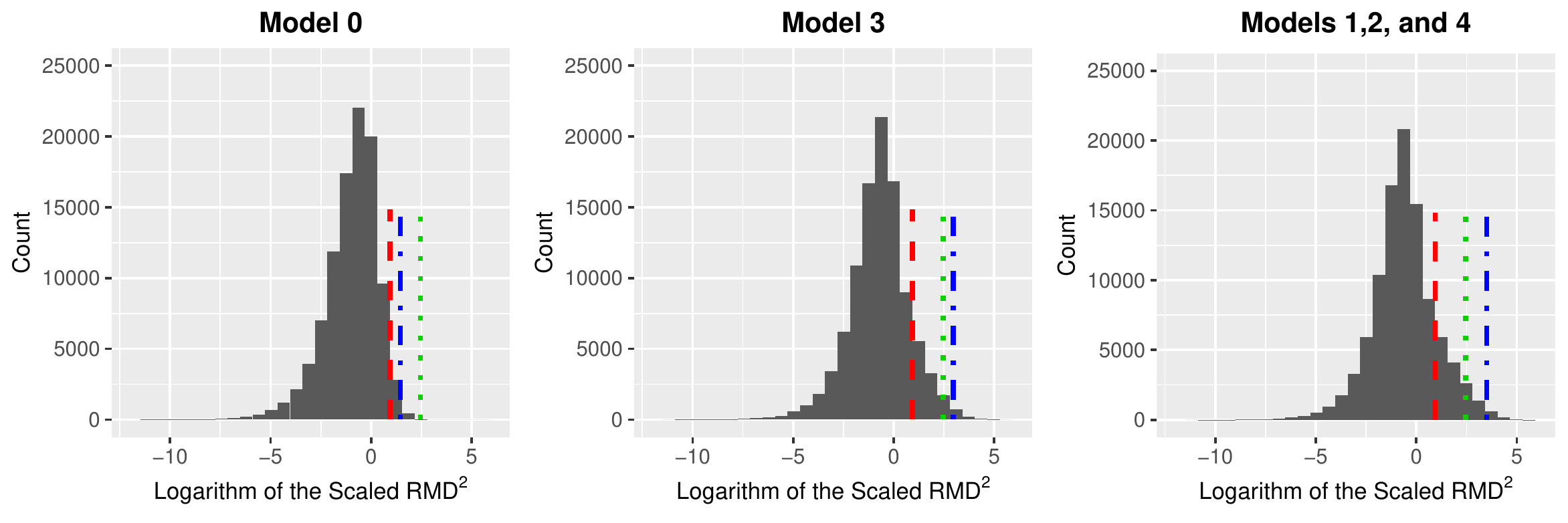}
		\caption{Histograms of the logarithm of the scaled ${\rm RMD}^2$ calculated from the models with different covariance functions. The red dashed line: the 0.993 quantile of the $\chi^2$ distribution with degree of freedom 2. The green dotted line: the 0.993 quantile of the $F$-distribution with simulated parameters. The blue dash-dot line: the 0.993 empirical quantile of the data. The quantile of the $F$-distribution is more accurate as the cutoff value.}
		\label{models}
	\end{center}
\end{figure}

For Model 0, the effective spatial range (the lag corresponding to 5$\%$ correlation) is about 0.003 and the $F$-distribution approximates the tail distribution of scaled ${\rm RMD}^2$ quite well; see the first plot of Figure \ref{models}.
However, Models 1-4 all have long effective spatial ranges of their covariance functions. Specifically, this range is 3 for Models 1, 2, and 4, and 0.9 for Model 3, which are relatively long for the $[0, 1]$ design interval. As a result, the distribution of ${({\rm MO}_{T_k,n},{\rm VO}_{T_k,n})}^{\rm T}$ deviates a little from the normal distribution, so the tail distribution of the scaled ${\rm RMD}^2$ is not as well approximated as that of Model 0; see the second and third plots of Figure \ref{models}. 
Nevertheless, our simulation study shows that the cutoff value from the $F$-distribution still leads to reasonable outlier detection results for such settings. Simulation results for models with short effective spatial ranges are provided in an online supplementary document.

\begin{sidewaystable}
	%\begin{table}
	\small
	\centering
	\caption{\small Mean and standard deviation (in parentheses) of the percentage of correctly and falsely detected outliers over 500 simulation runs with four univariate models. \textbf{Bold} font is used to highlight the worst performance on both measurements for each setting. Int.Sqe: integrated squared error method; Rob.Mah: robust Mahalanobis distance; Out.Grm: adjusted outliergram; Dir.Out: method based-on directional outlyingness.}
	\label{Table_uni}
	\begin{tabular}{lllllllll}
		\hline
		\multirow{2}{*}{Method} &\multicolumn{2}{c}{Model 1} &\multicolumn{2}{c}{Model 2} &\multicolumn{2}{c}{Model 3} &\multicolumn{2}{c}{Model 4}\\
		\cline{2-3} \cline{4-5} \cline{6-7}\cline{8-9}
		&$p_c$&$p_f$&$p_c$&$p_f$&$p_c$&$p_f$&$p_c$&$p_f$\\
		\hline
		$\epsilon=0$&&&&&&&&\\
		Int.Sqe   &  -& \textbf{7.1} (2.7)   & - & \textbf{7.1} (2.7)   & - & \textbf{4.1} (2.0)   & - & \textbf{7.1} (2.7)   \\
		%Dep.Trim &  -& 00.0(00.0)  & - & 00.0 (00.0) & - & 00.0 (00.0) & - & 0.0 (0.0)   \\
		Rob.Mah   &  -& 1.5 (1.3)   & - & 1.5 (1.3)   & - & 1.5 (1.3)   & - & 1.5 (1.3)   \\
		%Adj.FBP  &  -& 00.0(00.0)  & - & 00.0 (00.0) & - & 00.0 (00.0) & - & 0.0 (0.0)   \\
		Out.Grm   &  -& 1.2 (1.2)   & - & 1.2 (1.2)   & - & 1.0 (1.3)   & - & 1.2 (1.2)   \\
		Dir.Out   &  -& 1.9 (1.6)   & - & 1.9 (1.6)   & - & 1.0 (1.1)   & - & 1.9 (1.6)   \\
		\hline
		$\epsilon=0.1$&&&&&&&&\\
		Int.Sqe   &   \textbf{37.6} (30.1) & \textbf{6.1} (2.6)   & 100 (0.0)   & \textbf{6.3} (2.7)   & \textbf{55.3} (36.8) & \textbf{3.5} (1.9)   & 100 (0.0)   & \textbf{5.8} (2.4) \\
		%Dep.Trim &   00.0 (00.0) & 00.0 (00.0) & 00.0 (00.0) & 00.0(00.0)  & 00.0 (00.0) & 00.0 (00.0) & 00.0 (00.0) & 00.0 (00.0) \\
		Rob.Mah   &   100 (0.0)   & 0.4 (0.7)   & \textbf{25.4} (15.5) & 0.8 (1.1)   & 98.9 (4.8)  & 0.4 (0.8)   & \textbf{11.8} (10.1) & 1.0 (1.3) \\
		%Adj.FBP  &   00.0(00.0)  & 00.0 (00.0) & 00.0 (00.0) & 00.0 (00.0) & 00.0 (00.0) & 00.0 (00.0) & 00.0 (00.0) & 00.0 (00.0) \\
		Out.Grm   &   100 (0.0)   & 2.1 (2.1)   & 67.8 (20.2) & 3.6 (3.2)   & 82.3 (18.7) & 1.9 (1.7)   & 85.1 (16.3) & 3.3 (2.2) \\
		Dir.Out   &   99.6 (2.1)  & 1.1 (1.2)   & 100 (0.0)   & 1.0 (1.2)   & 99.8 (1.8)  & 0.6 (0.9)   & 98.6 (4.1)  & 1.0 (1.2) \\
		\hline
		$\epsilon=0.2$&&&&&&&&\\
		Int.Sqe   &   \textbf{29.9} (24.9) & 5.5 (2.8)   & 100 (0.0)   & 5.1 (2.6)   & 43.0 (33.5) & 3.1 (1.9)   & 100 (0.0)   & 4.8 (2.4) \\
		%Dep.Trim &   00.0 (00.0) & 00.0 (00.0) & 00.0 (00.0) & 00.0 (00.0) & 00.0 (00.0) & 00.0 (00.0) & 00.0 (00.0) & 00.0 (00.0) \\
		Rob.Mah   &   100 (0.0)   & 0.0 (0.2)   & \textbf{15.4} (10.2) & 0.3 (0.7)   & 91.5 (12.6) & 0.0 (0.1)   & \textbf{9.2}  (6.8)  & 0.7 (1.0) \\
		%Adj.FBP  &   00.0 (00.0) & 00.0 (00.0) & 00.0 (00.0) & 00.0 (00.0) & 00.0 (00.0) & 00.0 (00.0) & 00.0 (00.0) & 00.0 (00.0) \\
		Out.Grm   &   100 (0.2)   & \textbf{14.7} (9.9)  & 64.5 (17.4) & \textbf{7.9} (5.3)   & \textbf{29.1} (16.5) & 1.2 (1.6)   & 85.1 (14.9) & \textbf{5.1} (2.9) \\
		Dir.Out   &   99.3 (2.9)  & 0.4 (0.7)   & 100 (0.0)   & 0.3 (0.6)   & 94.6 (6.8)  & 0.2 (0.5)   & 89.5 (12.9) & 0.3 (0.7) \\
		\hline
	\end{tabular}
	%\end{table}
\end{sidewaystable}

When the model is clean ($\epsilon=0$), Int.Sqe suffers from high false detection rates compared with the other methods.
Each of the three methods performs poorly for either $p_c$ or $p_f$ with the contamination models.
In particular, Int.Sqe fails to handle Models 1 and 3, Rob.Mah performs poorly with Models 2 and 4 and Out.Grm produces the highest false detection rate when $\epsilon=0.2$.
It is also not robust to the rising contamination level (see Model 3, $p_c$ for $\epsilon=0.1$ and $0.2$).
In contrast, Dir.Out generates quite stable and satisfying correct detection rates with a low false detection rate.
All these results indicate that our proposed method based on directional outlyingness can detect various types of outliers.

\subsection{Multivariate Functional Data}
For multivariate functional data, we consider WMBD \citep{ieva2013depth} and MSBD \citep{lopez2014simplicial} as two competitors.
WMBD is constructed as a weighted average of the marginal modified band depth of each variable of the multivariate functional data.
The adjusted functional boxplot \citep{sun2012adjusted} based on WMBD is applied to detect marginal outliers and the final result is the union of these marginal outliers.
MSBD is defined as the mean of point-wise simplicial depths across the whole design interval. We follow the outlier detection procedure based on MSBD from \citet{lopez2014simplicial}.
For our method, we apply it in two ways: 1) treating each variable as a univariate curve, detecting marginal outliers and combining them together as the final result (Mar.Dir.Out); 2) taking bivariate curves as a whole and detecting the overall outliers (Tot.Dir.Out).

We consider six groups of bivariate curves containing outliers with different types and levels of outlyingness. Following the setting in \citet{lopez2014simplicial}, we define the main model as $\mathbf{X}(t)=\mathbf{e}(t)$, where $\mathbf{e}(t)=\{{\rm e}_1(t), {\rm e}_2(t)\}^{\rm T}$ is a bivariate Gaussian process with zero mean and cross-covariance function \citep{gneiting2010matern,apanasovich2012valid}:
$$C_{ij}(s,t)=\rho_{ij}\sigma_{i}\sigma_{j}\mathcal{M}(|s-t|;\nu_{ij},\alpha_{ij}), \quad i,j=1,2,$$
where $\rho_{12}$ is the correlation between $X_1(t)$ and $X_2(t)$, $\rho_{11}=\rho_{22}=1$, $\sigma_i^2$ is the marginal variance and $\mathcal{M}(h;\nu,\alpha)=2^{1-\nu}\Gamma(\nu)^{-1}\left(\alpha|h|\right)^{\nu}\mathcal{K}_\nu(\alpha|h|)$ with $|h|=|s-t|$, is the Mat{\'e}rn class \citep{matern1960spatial} where $\mathcal{K}_\nu$ is a modified Bessel function of the second kind, $\nu>0$ is a smoothness parameter, and $\alpha>0$ is a range parameter. Throughout the simulation, we choose the following parameters for the bivariate Mat{\'e}rn cross-covariance function: $\sigma_1=\sigma_2=1$, $\alpha_{11}=0.02$, $\alpha_{22}=0.01$, $\alpha_{12}=0.016$, $\nu_{11}=1.2$, $\nu_{22}=0.6$, $\nu_{12}=1$ and $\rho_{12}=0.6$. The six models investigated are as follows:
\begin{itemize}[noitemsep]
	\item Model 5 (uncontaminated model). Main model: $\mathbf{X}(t)=\mathbf{e}(t)$.
	\item Model 6 (consistent outlier). Main model: $\mathbf{X}(t)=\mathbf{e}(t)$ and contamination model: $X_i(t)=4{\rm e}_i(t)$, $i=1,2$.
	\item Model 7 (isolated outlier). Main model: $\mathbf{X}(t)=\mathbf{e}(t)$ and contamination model: $X_i(t)={\rm e}_i(t)(1+11I_{\{T\le t\le T+0.1\}})$, $i=1,2$.
	\item Model 8 (weak consistent outlier). Main model: $\mathbf{X}(t)=\mathbf{e}(t)$ and contamination model: $X_1(t)=1.7{\rm e}_1(t)$ and $X_2(t)=1.5{\rm e}_2(t)$. This is a weak version of Model 6.
	\item Model 9 (weak isolated outlier). Main model: $\mathbf{X}(t)=\mathbf{e}(t)$ and contamination model: $X_i(t)={\rm e}_i(t)(1+4I_{\{T\le t\le T+0.1\}})$, $i=1,2$. This is a weak version of Model 7.
	\item Model 10 (shape outlier). Main model: $X_1(t)={\rm e}_1(t)+U_{11}\cos(4\pi x)$ and $X_2(t)={\rm e}_2(t)+U_{12}\sin(4\pi x)$, where $U_{11}$ and $U_{12}$ are independent and follow a uniform distribution on $[2,3]$; contamination model: $X_1(t)={\rm e}_1(t)+U_{21}\cos(4\pi x)$ and $X_2(t)={\rm e}_2(t)+U_{22}\sin(4\pi x)$, where $U_{21}$ and $U_{22}$ are independent and follow a uniform distribution on $[3.2,3.5]$.
\end{itemize}

\begin{table}[t!]
	\scriptsize
	\centering
	\caption{\small Mean and standard deviation (in parentheses) of the percentage of correctly and falsely recognized outliers over 500 simulation runs with six bivariate models. \textbf{Bold} font is used to highlight the worst performance on both measurements for each setting. AdjFB.WMBD: adjusted functional boxplot using weighted modified band depth; AdjFB.MSBD: adjusted functional boxplot using modified simplicial depth; Mar.Dir.Out: detect outliers marginally with Dir.Out and combine them together as the final result; Tot.Dir.Out: detect outliers jointly with Dir.Out.}
	\label{Table_biv}
	\begin{tabular}{lllllll}
		\hline
		\multirow{2}{*}{Method} &\multicolumn{2}{c}{Model 5} &\multicolumn{2}{c}{Model 6} &\multicolumn{2}{c}{Model 7} \\
		\cline{2-3} \cline{4-5} \cline{6-7}
		&$p_c$&$p_f$&$p_c$&$p_f$&$p_c$&$p_f$\\
		\hline
		AdjFB.WMBD    &  -    & 0.0 (0.0)    & \textbf{31.7} (15.0) & 0.0 (0.0)   & 83.8 (12.4)& 0.0 (0.0)    \\
		AdjFB.MSBD    &  -    & 0.0 (0.0)    & 63.0 (15.6) & 0.0 (0.0)   & \textbf{74.1} (16.3)& 0.0 (0.0)    \\
		Mar.Dir.Out   &   -   & \textbf{2.0} (0.5)    & 100 (0.0)   & 0.0 (0.0)   & 100 (0.0)  & \textbf{0.0} (0.2)    \\
		Tot.Dir.Out   &   -   & 0.0 (0.2)    & 100 (0.0)   & \textbf{0.0} (0.2)   & 100 (0.0)  & \textbf{0.0} (0.2)    \\
		\hline
		\multirow{2}{*}{Method} &\multicolumn{2}{c}{Model 8} &\multicolumn{2}{c}{Model 9} &\multicolumn{2}{c}{Model 10} \\
		\cline{2-3} \cline{4-5} \cline{6-7}
		&$p_c$&$p_f$&$p_c$&$p_f$&$p_c$&$p_f$\\
		\hline
		AdjFB.WMBD    &   \textbf{0.0} (0.0)    & 0.0 (0.0)     & \textbf{0.0} (0.0)   & 0.0 (0.0)    & \textbf{0.0} (0.0)    & 0.0 (0.0)     \\
		AdjFB.MSBD    &   \textbf{0.0} (0.0)    & 0.0 (0.0)     & \textbf{0.0} (0.0)   & 0.0 (0.0)    & \textbf{0.0} (0.0)    & 0.0 (0.0)     \\
		Mar.Dir.Out   &   59.6 (17.7)  & 0.0 (0.0)     & 81.6 (12.7) & 0.0 (0.0)    & 66.1 (16.6)  & 0.0 (0.1) \\
		Tot.Dir.Out   &   83.2 (13.0)   & \textbf{0.0} (0.1)     & 93.7 (7.6)  & \textbf{0.0} (0.1)    & 81.8 (13.4)  & \textbf{0.0} (0.1) \\
		\hline
	\end{tabular}
\end{table}

We set the contamination level in the contamination models to $\epsilon=0.1$. For each case, we generate 100 bivariate curves on 50 equally spaced time points on $[0,1]$. The same $h=[0.75n]$ is employed.
The correct detection rate, $p_c$, and false detection rate, $p_f$, are calculated and reported in Table \ref{Table_biv}.
For all six models, the four methods are quite satisfactory by generating small $p_f$.
For Models 6 and 7, the proposed two methods perfectly detect all the outliers, a result much better than the results from the two competitors.
For Models 8, 9 and 10, WMBD and MSBD fail to detect any outlier due to the conservativeness of the functional boxplot. Furthermore, for these three models, Tot.Dir.Out has significant improvement over Mar.Dir.Out, which suggests that it is indeed better to treat the bivariate curves as a whole when detecting outliers.

%
%\begin{figure}
%\begin{center}
%\includegraphics[width=15cm,height=15cm]{illust1}\\
%\label{first_derivative_lowerbias}
%\end{center}
%\end{figure}

\section{Data Applications}
\subsection{Spanish Weather Data}

Besides simulations, we test our proposed framework with two datasets.
The first dataset is spanish weather data from the R package \emph{fda.usc}.
This dataset contains geographic information of 73 weather stations in Spain, from where 
average daily temperature and daily log precipitation for the period 1980-2009 were recorded. 
The data are discretely observed and hence smoothed with 11 order-4 B-spline basis functions. 
According to our experience, the plot of $(\mathbf{MO}^{\rm T},{\rm VO})^{\rm T}$ is quite robust to the amount of smoothing (see Figure \ref{robust}) and to the choice of smoothing methods (see Figure S1 of the supplementary material.)

\begin{figure}[b!]
	\begin{center}
		\includegraphics[width=12cm,height=8cm]{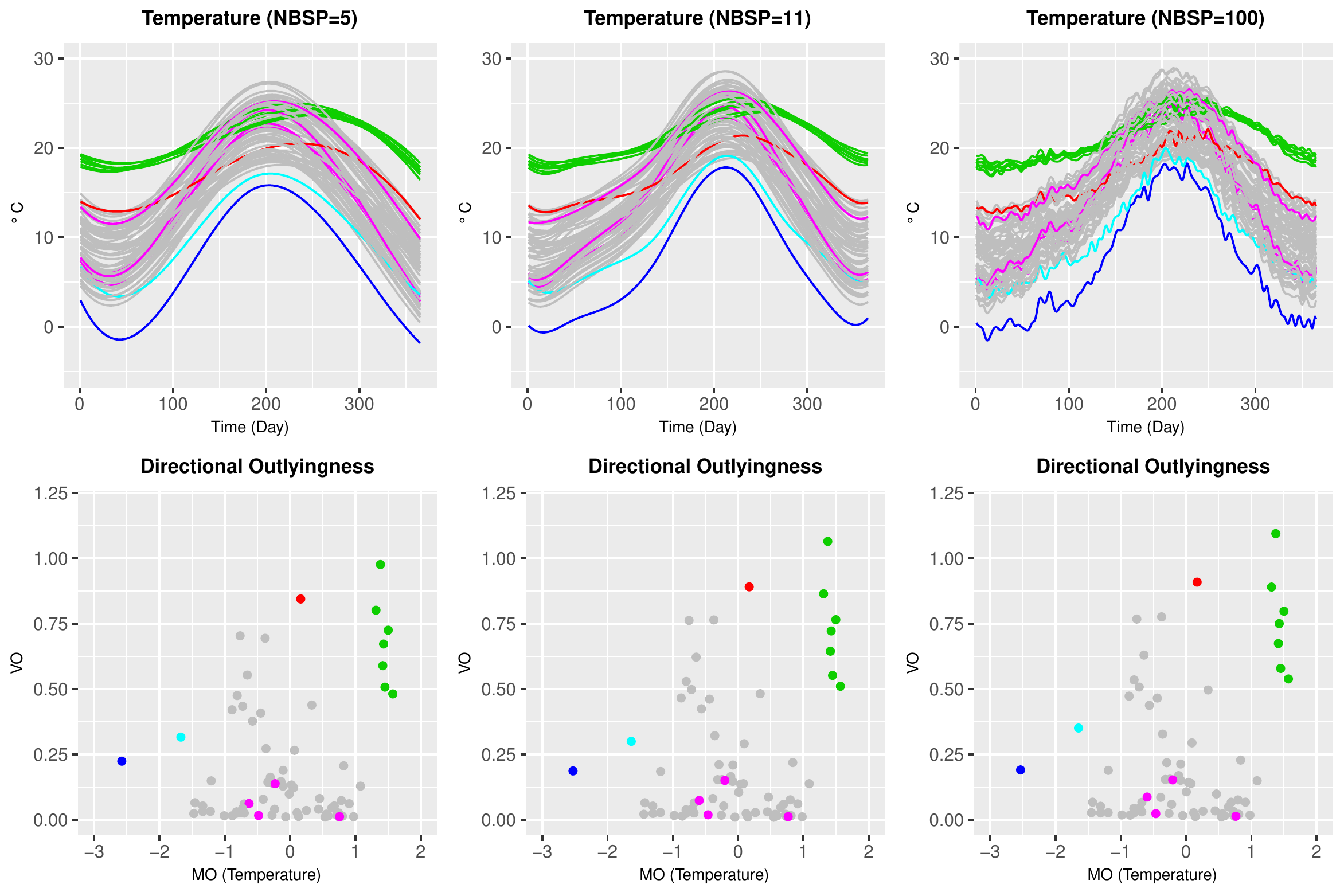}
		\caption{Top panel: temperature curves smoothed with different numbers of B-spline basis functions (NBSP).  Bottom panel: directional outlyingness plot corresponding to the fitted curves.}
		\label{robust}
	\end{center}
\end{figure}

\begin{figure}[t!]
	\begin{center}
		\includegraphics[width=12cm,height=12cm]{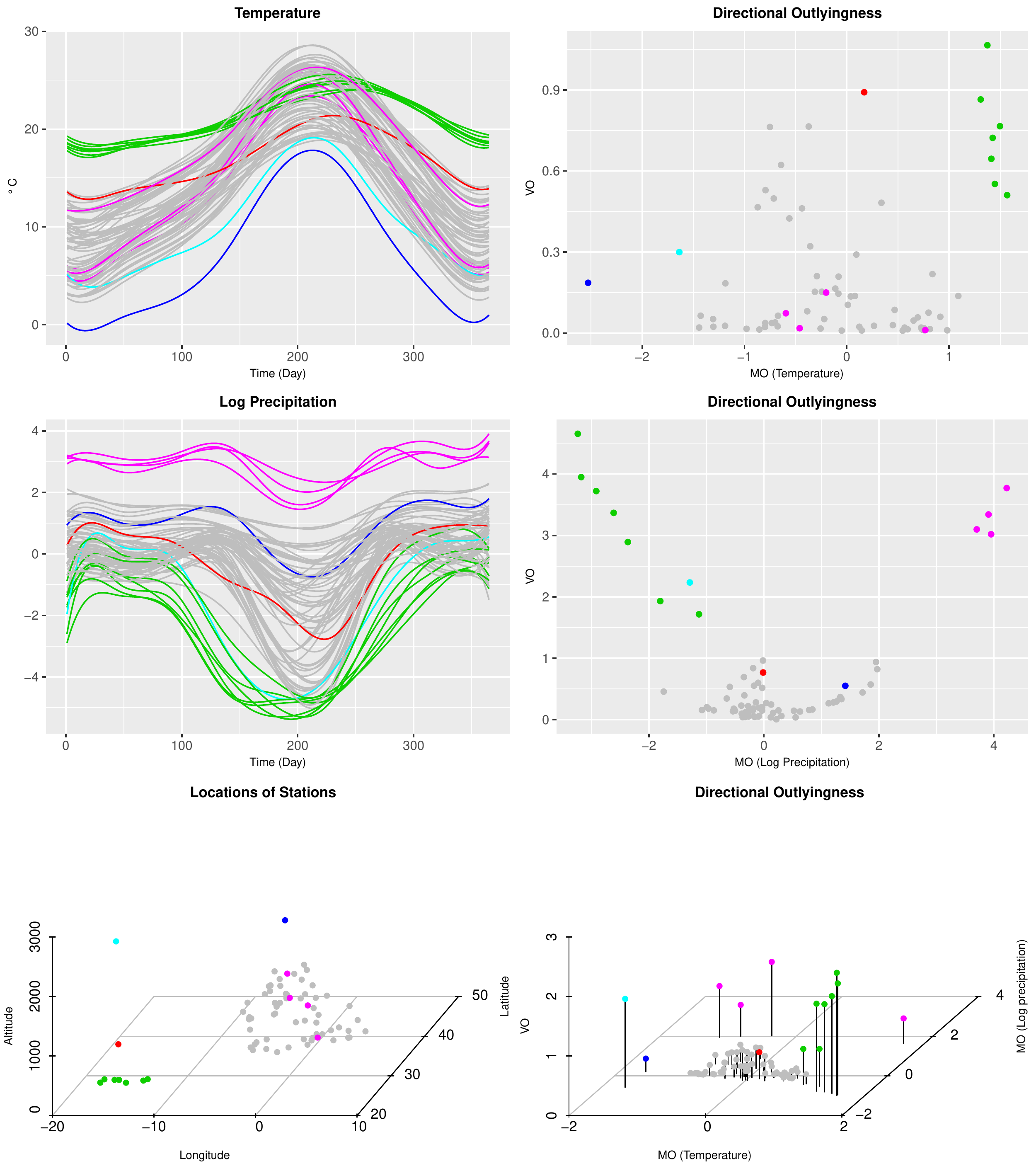}
		\caption{Top panel: temperature curves (left) and directional outlyingness plot (right); middle panel: log precipitation curves (left) and directional outlyingness plot (right); bottom panel: locations of stations (left) and directional outlyingness plot for bivariate curves (right). Colors represent different groups of outlying curves.}
		\label{example_visualiztion}
	\end{center}
\end{figure}

We separate the stations into six groups mainly with respect to their geographic locations, which have great influence on the weather and check if the proposed method can also distinguish such groups based on the weather records. Specifically, we calculate the functional directional outlyingness for the temperature curves, the log precipitation curves, and the bivariate curves combining two types of data together, respectively.
We then illustrate the results for the three cases (Figure \ref{example_visualiztion}).
For the univariate cases, we see clear patterns from the directional outlyingness plot corresponding with the magnitudes and shapes of the curves.
For instance, the seven green and two blue temperature curves are mapped as points isolated from the main cluster (Figure \ref{example_visualiztion}, top) and the green and purple log precipitation curves are also well distinguished in the directional outlyingness plot (Figure \ref{example_visualiztion}, middle).
For the bivariate case, all the six groups of stations are clearly located in the directional outlyingness plot.
These results demonstrate that the functional directional outlyingness provides an effective way to visualize the centrality of functional data. 

\begin{figure}[b!]
	\begin{center}
		\includegraphics[width=8cm,height=7.2cm]{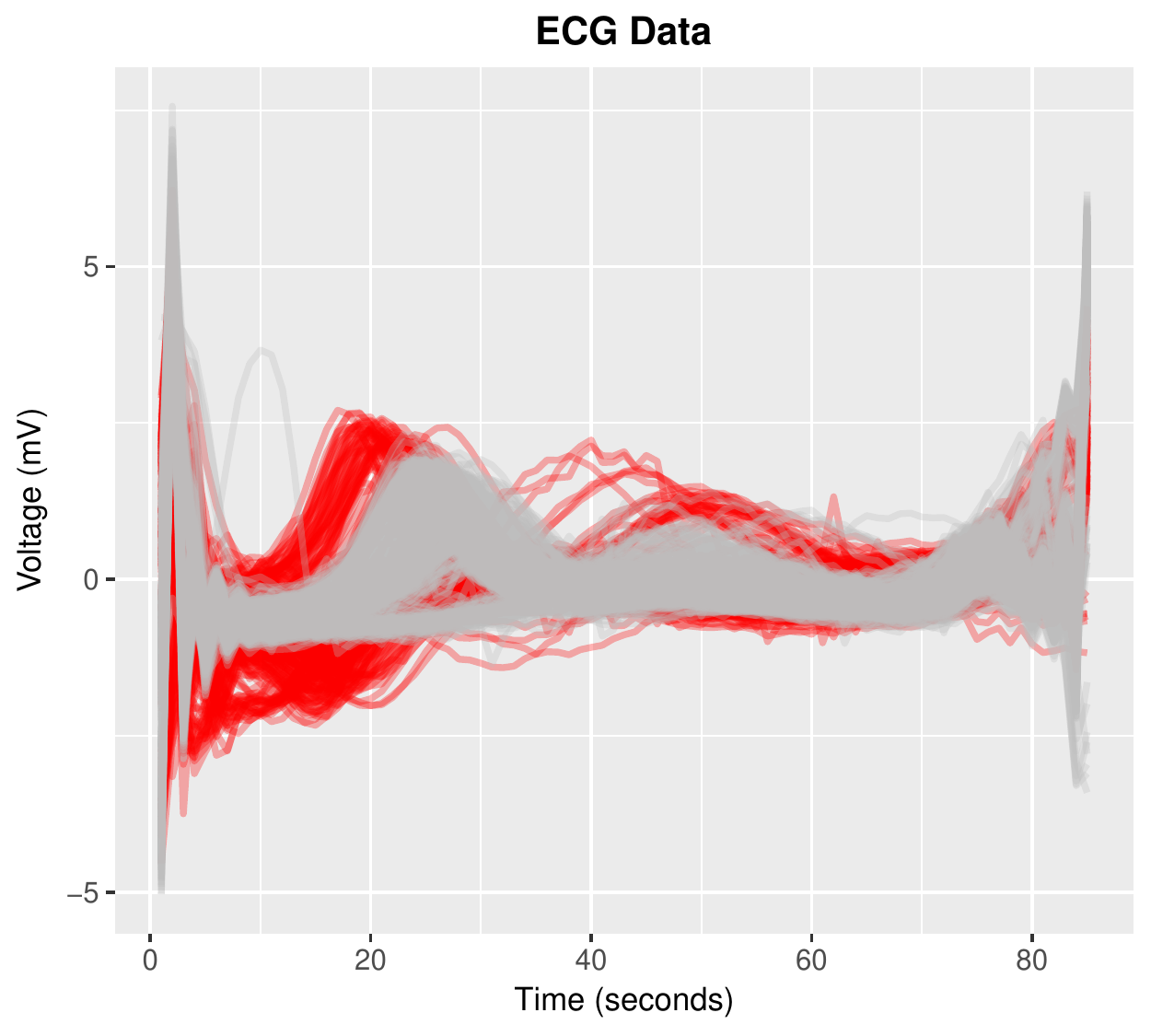}\\
		\caption{The ECG data: normal curves (grey); abnormal curves (red).}
		\label{ECG_data}
	\end{center}
\end{figure}

%
%
%\begin{figure}[b!]
%\begin{center}
%\includegraphics[width=15cm,height=15cm]{example_test}
%\caption{Outlier detection results of four methods: integral squared error, outliergram, Robust Mahalanobis distance and directional outlyingness. Normal curves (grey) and detected outliers (red).}
%\label{example_test}
%\end{center}
%\end{figure}
%
%
%
%
%
%We also detect outliers from the temperature data using the four aforementioned methods in the simulation studies. For our method, we choose $h=[0.55n]$ for the sake of robustness.
%Based on the locations of the weather stations, we find that the green curves are observed from stations in the provinces of \emph{Santa Cruz de Tenerife} and \emph{Las Palmas}, which are islands off northwestern Africa and far away from mainland Spain.
%The annual temperatures from these weather stations demonstrate very different patterns from those recorded in mainland Spain.
%Thus, we can treat the observations from stations on islands as potential outliers.
%The test results are presented in Figure \ref{example_test}.
%Only our method and Int.Sqe find the abnormality of the potential outliers, whereas the other two methods find nearly nothing but one curve by Out.Grm.
%Int.Seq is quite sensitive since it detects 25 (out of 73) temperature curves as outliers, which is consistent with its high false detection rate as shown in the simulation studies.
%Dir.Out detects only four more curves as outliers besides seven obvious outlying curves, which is the most reasonable result among the four methods.

\subsection{ECG Data}
The second dataset consists of electrocardiogram (ECG) data, which is a measure of how the electrical activity of the heart changes over time as action potentials propagate throughout the heart during each cardiac cycle.
ECG data are used to conduct heartbeat classification to diagnose abnormalities in heart activities.
We obtained the data from the MIT-BIH Arrhythmia Database \citep{goldberger2000physiobank} and \citet{wei2006semi}.
The data were annotated by cardiologists and a label of \emph{normal} or \emph{abnormal} was assigned to each data record; see Figure \ref{ECG_data}.

All the data records were normalized and rescaled to have lengths of 85.
We took all the training samples in \citet{wei2006semi} including 208 abnormal and 602 normal cases.
Only observations on time points 6 to 80 were considered to avoid boundary effects.
We detected abnormal curves (outliers) with six methods including the four methods used in Sections 5.1 and 6.1, the functional directional outlyingness method based on both the mean function and the first-order derivative function (Dir.Out1), and the functional directional outlyingness method based on the mean function, the first- and second-order derivative functions (Dir.Out2).
We randomly selected 400 curves from the total 810 observations, controlling the contamination level (CL) at $0, 5\%, 10\%, 15\%, 20\%$ and $25\%$, respectively.
The CL of the raw data is $25.7\%$, so we set the maximum CL as $25\%$.
The 400 curves were selected in this way: when CL is $10\%$, for example, we randomly chose 40 from 208 abnormal curves and 360 from 602 normal curves.
For each CL, we calculated the detection rates, $p_c$ and $p_f$, for each method.
We repeated this procedure for 50 times and report the average rates in Table \ref{Table_example2}.

\begin{table}[b!]
	\small
	\centering
	\caption{\small A comparison of the outlier detection performances of the six methods applied to the ECG data of different contamination levels (CL). Int.Sqe: integrated squared error method; Rob.Mah: robust Mahalanobis distance; Out.Grm: adjusted outliergram; Dir.Out, Dir.Out1, and Dir.Out2: methods based-on directional outlyingness.}
	\label{Table_example2}
	\begin{tabular}{rrrrrrrr}
		\toprule[1pt]
		CL    & Rate         &Int.Sqe &Rob.Mah&Out.Grm&Dir.Out&Dir.Out1&Dir.Out2  \\
		\multirow{2}{*}{0}   & $p_c$       &   -    & -     & -     & -     & -     &  -    \\
		& $p_f$       &   16.7 & 13.1  &  5.3  & 17.3  & 18.0  &  20.3 \\       \hline
		\multirow{2}{*}{5\%} & $p_c$       &   96.0 & 48.7  & 67.3  & 90.1  & 99.3  &  100  \\
		& $p_f$       &   16.6 & 13.0  & 4.9   & 15.6  & 15.2  &  17.3 \\       \hline
		\multirow{2}{*}{10\%}& $p_c$       &   93.3 & 39.7  & 56.3  & 86.9  & 98.2  &  99.5 \\
		& $p_f$       &   15.7 & 12.4  & 4.0   & 12.8  & 12.4  &  12.3 \\       \hline
		\multirow{2}{*}{15\%}& $p_c$       &   90.9 & 28.6  & 46.0  & 81.7  & 96.4  &  99.2 \\
		& $p_f$       &   15.3 & 11.2  & 3.2   & 10.0  & 11.0  &  10.4 \\       \hline
		\multirow{2}{*}{20\%}& $p_c$       &   85.3 & 20.0  & 33.8  & 74.1  & 93.1  &  97.5 \\
		& $p_f$       &   14.2 &  8.3  &  2.3  & 7.3   &  8.4  &   7.2 \\      \hline
		\multirow{2}{*}{25\%}& $p_c$       &   76.6 & 7.1   & 19.6  & 60.4  & 86.1  &  92.9 \\
		& $p_f$       &   12.9 & 1.6   &  1.7  & 4.2   &  5.8  &   5.8  \\
		\bottomrule[1pt]
	\end{tabular}
\end{table}

First of all, Out.Grm and Rob.Mah are not acceptable due to their poor correct detection rate when ${\rm CL}>0$.
When ${\rm CL}=0$, Int.Sqe and Dir.Out all suffer from high false detection rates because even the normal curves (grey ones in Figure \ref{ECG_data}) demonstrate significantly different patterns (the segment between times 20 and 40 for example).
That is to say, although these people are diagnosed as normal, some individual differences of heart activities still exist among them.
As CL increases, Dir.Out type methods reduce the false detection rate more rapidly than Int.Sqe.
For the correct detection rate, Int.Sqe is better than Dir.Out but worse than Dir.Out1 and Dir.Out2.
These results indicate that it is better to consider the derivative functions together with the mean function in practical implementations of the outlier detection method based on directional outlyingness.

We also compared the computational speeds of the six methods by applying them to the whole 810 curves. The results are reported in Table \ref{Table_Time}. Int.Sqe is shown to take the longest time, which is about 240 times slower than the fastest method, Dir.Out.
After increasing the dimension of curves, Dir.Out1 and Dir.Out2 take longer time than Dir.Out since the calculation of depth for multivariate data is more time-consuming than for univariate data.
Nevertheless, Dir.Out1 and Dir.Out2 are still much faster than Int.Sqe.

\begin{table}[t!]
	\small
	\centering
	\caption{\small A comparison of the computational speeds of the six methods applied to the ECG data.}
	\label{Table_Time}
	\begin{tabular}{crrrrrr}
		\toprule[1pt]
		&Int.Sqe&Out.Grm&Rob.Mah&Dir.Out&Dir.Out1&Dir.Out2  \\
		\hline
		Time (secs) &  144.7 & 20.7  & 3.5   & 0.6   & 15.4  &  17.7  \\
		\bottomrule[1pt]
	\end{tabular}
\end{table}

\vskip 6pt
\section{Discussion}

Unlike with point-wise data, the direction of outlyingness is crucial to describing the centrality of multivariate functional data.
Motivated by this, we proposed a new framework of directional outlyingness for multivariate functional data by taking the direction of outlyingness into consideration.
Compared with classical functional depth, functional directional outlyingness reveals its superiority by naturally decomposing the total outlyingness into two parts: magnitude outlyingness and shape outlyingness, and by visualizing the centrality of a group of curves. 
We created an outlier detection criterion, applicable to both univariate and multivariate curves, based on functional directional outlyingness. 
This criterion outperformed competing methods in simulation studies.
Moreover, all methods proposed here can be readily extended to the case where the design area is two-dimensional, corresponding to image data \citep{genton2014surface}.

Most classical functional depths map a curve (both univariate and multivariate) to a univariate depth curve (a sequence of nonnegative scalar depth, discretely).
Most existing methods can be derived from various ways of dealing with such a depth curve.
For example, by taking the weighted average across the design interval, we obtain ID, MBD, MSBD, MHD and MFHD; by taking the smallest value of the depth curve, we obtain BD, HD and SBD; by stochastically ordering the left tail of the distributions of these point-wise depths, we obtain the extremal depth.
However, the functional directional outlyingness maps each curve to an outlyingness curve (or a sequence of vectors, discretely) of the same dimension.
This preservation of dimension provides much more information and flexibility for functional data visualization and outlier detection.
In the current paper, we made use of this sequence of vectors in two different ways: taking their weighted average leads to \textbf{MO} and calculating their variance leads to VO.
Other ways of handling this sequence are to be further explored, so that different measures for the centrality of functional data can be obtained for various purposes.

Another good feature of functional directional outlyingness is the outlyingness decomposition, which allows us to demonstrate the centrality of the curves and also the type of outliers more accurately.
Similar ideas have been proposed based on classical point-wise depth in the literature. \citet{hubert2015multivariate} proposed multivariate functional skew-adjusted projection depth (MFSPD) to be a weighted average of point-wise, skew-adjusted projection depth (SPD). They used $1-{\rm MFSPD}$ as the overall depth and treated the difference between the harmonic mean and the arithmetic mean of the point-wise SPD as the shape depth.
\citet{hubert2016finding} measured overall outlyingness with the weighted average of point-wise adjusted outlyingness (AO) and the shape outlyingness with the scaled standard deviation of AO.
However, their decomposition is less natural in that the two components are dependent on each other, and also their measurements of the shape depth (outlyingness) is less accurate compared with our VO due to the lack of the direction of outlyingness, a fact also discussed by \citet{discussion2015narisetty}.

%
%Most existing method define the functional depth as a weighted average of the point-wise depth.  While, there are also some other way to deal with these point-wise depth. The extremal depth treated a group of point-wise depth as a distribution and they rank the curve with respect to the left tail of each empirical distribution. This inspires the possibility of similar way to deal with the directional point-wise depth we get.
%
%How to better measure the shape variation and magnitude variation more accurately?? Variogram of point-wise depth???
%
%Depth decomposition is actually achieved by taking the average of point-wise depth and the variation of them. If one do the same thing to the classical point-wise depth, a similar decomposition is accessible but is not as reasonable as the current one.  Some existing work by ?? and ??.

%Further research is needed on the classification of functional data based on directional outlyingness or the outlying matrix.

\vskip 6pt
\section*{Appendix}

\noindent
\emph{Proof of Theorem \ref{Theorem 1}.} We provide a proof of the first property. Proofs of the other two properties can be directly derived using definition (\ref{Def1}) and the properties of $d\left(\mathbf{X}(t),F_{\mathbf{X}(t)}\right)$. We have $d\left(\mathbf{A}\mathbf{X}(t)+\mathbf{b},F_{\mathbf{A}\mathbf{X}(t)+\mathbf{b}}\right)={d}\left(\mathbf{X}(t),F_{\mathbf{X}(t)}\right)$ and
$d\left(\mathbf{A}\mathbf{Z}(t)+\mathbf{b},F_{\mathbf{A}\mathbf{X}(t)+\mathbf{b}}\right)={d}\left(\mathbf{Z}(t),F_{\mathbf{X}(t)}\right)$ based-on the affine invariance of $d\left(\mathbf{X}(t),F_{\mathbf{X}(t)}\right)$.
Hence, $\|\mathbf{O}\left(\mathbf{A}\mathbf{X}(t)+\mathbf{b},F_{\mathbf{A}\mathbf{X}(t)+\mathbf{b}}\right)\|={\|\mathbf{O}\left(\mathbf{X}(t),F_{\mathbf{X}(t)}\right)\|}$. For the direction of this outlyingness, we have
\begin{eqnarray*}
	\mathbf{v^*}(t)&=&{\mathbf{A}\mathbf{X}(t)-\mathbf{A}\mathbf{Z}(t)\over \|\mathbf{A}\mathbf{X}(t)-\mathbf{A}\mathbf{Z}(t)\|}={\mathbf{A}\|\mathbf{X}(t)-\mathbf{Z}(t)\| \over \|\mathbf{A}\mathbf{X}(t)-\mathbf{A}\mathbf{Z}(t)\|}\cdot{\mathbf{X}(t)-\mathbf{Z}(t)\over \|\mathbf{X}(t)-\mathbf{Z}(t)\|}=\mathbf{U}\cdot\mathbf{v}(t).
\end{eqnarray*}
By the definition of $\mathbf{v^*}(t)$, we have
\begin{eqnarray*}
	\mathbf{O}\left(\mathbf{A}\mathbf{X}(t)+\mathbf{b},F_{\mathbf{A}\mathbf{X}(t)+\mathbf{b}}\right)&=&\mathbf{v^*}(t)\|\mathbf{O}\left(\mathbf{A}\mathbf{X}(t)+\mathbf{b},F_{\mathbf{A}\mathbf{X}(t)+\mathbf{b}}\right)\|\\
	&=&\mathbf{U}\cdot\mathbf{v}(t)\|\mathbf{O}\left(\mathbf{X}(t),F_{\mathbf{X}(t)}\right)\|=\mathbf{U}\cdot\mathbf{O}\left(\mathbf{X}(t),F_{\mathbf{X}(t)}\right).
\end{eqnarray*}
This completes the proof.  \hfill $\Box$

\vskip 5pt
\noindent
\emph{Proof of Theorem \ref{Consistency of pointwise depth}.} With the consistency of the median, we get $\mathbf{v}_n(t)\to \mathbf{v}(t)$. Then, combined with the convergence of classical depth, we get the consistency of directional outlyingness. \hfill $\Box$

\vskip 5pt
\noindent
\emph{Proof of Theorem \ref{Theorem 2}.}
We decompose the total outlyingness into two parts: magnitude outlyingness and shape outlyingness:
\begin{eqnarray*}
	{\rm FO}(\mathbf{X},F_{\mathbf{X}})&=&\int_{\mathcal{I}} \|\mathbf{O}(\mathbf{X}(t),F_{\mathbf{X}(t)})\|^2w(t){\rm d}t\\
	&=&\int_{\mathcal{I}} \|\mathbf{O}(\mathbf{X}(t),F_{\mathbf{X}(t)})-\mathbf{MO}(\mathbf{X},F_{\mathbf{X}})+\mathbf{MO}(\mathbf{X},F_{\mathbf{X}})\|^2w(t){\rm d}t\\
	&=&\int_{\mathcal{I}} \|\mathbf{O}(\mathbf{X}(t),F_{\mathbf{X}(t)})-\mathbf{MO}(\mathbf{X},F_{\mathbf{X}})\|^2w(t)dt+\|\mathbf{MO}(\mathbf{X},F_{\mathbf{X}})\|^2\\
	&=&{\rm VO}(\mathbf{X},F_{\mathbf{X}})+\|\mathbf{MO}(\mathbf{X},F_{\mathbf{X}})\|^2.
\end{eqnarray*}
\vskip -32pt
\hspace*{\fill}{$\Box$}

\noindent
\emph{Proof of Theorem \ref{Theorem 3}.}
\noindent
1(a) Since $g$ is a one-to-one transformation on the interval $\mathcal{I}$,
\begin{eqnarray*}
	{\rm FO}\left(\mathbf{X}_g,F_{\mathbf{X}_g}\right)&=&\int_{\mathcal{I}} \|\mathbf{O}\left(\mathbf{X}\{g(t)\},F_{\mathbf{X}\{g(t)\}}\right)\|^2w\{g(t)\}{\rm d}t\\
	&=&\int_{g^{-1}(\mathcal{I})} \|\mathbf{O}\left(\mathbf{X}(t),F_{\mathbf{X}(t)}\right)\|^2w(t){\rm d}t={\rm FO}\left(\mathbf{X},F_{\mathbf{X}}\right).
\end{eqnarray*}
Also, we have ${\rm FO}\left(\mathbf{X}_g,F_{\mathbf{X}_g}\right)={\rm FO}\left(\mathbf{T}(\mathbf{X}_g),F_{\mathbf{T}(\mathbf{X}_g)}\right)$ by the affine invariance of $\mathbf{O}(\mathbf{X}(t),F_{\mathbf{X}(t)}$.
Consequently, we proved that ${\rm FO}\left(\mathbf{T}(\mathbf{X}_g),F_{\mathbf{T}(\mathbf{X}_g)}\right)={\rm FO}\left(\mathbf{X},F_{\mathbf{X}}\right)$.

\noindent
1(b) By a similar procedure with 1(a), we can show
\begin{eqnarray}
\mathbf{MO}\left(\mathbf{T}(\mathbf{X}),F_{\mathbf{T}(\mathbf{X})}\right)=\mathbf{MO}\left(\mathbf{T}(\mathbf{X}_g),F_{\mathbf{T}(\mathbf{X}_g)}\right). \label{C1}
\end{eqnarray}
By the proof of Theorem \ref{Theorem 1}, we get
$$\mathbf{O}(\mathbf{T}(\mathbf{X}(t)),F_{\mathbf{T}(\mathbf{X}(t))})={f(t)\mathbf{A}_0\|\mathbf{X}(t)-\mathbf{Z}(t)\|\over \|f(t)\|\cdot\|\mathbf{A}_0\{\mathbf{X}(t)-\mathbf{Z}(t)\}\|}\mathbf{O}(\mathbf{X}(t),F_{\mathbf{X}(t)}),$$
and $\|\mathbf{X}(t)-\mathbf{Z}(t)\|=\|\mathbf{A}_0\{\mathbf{X}(t)-\mathbf{Z}(t)\}\|$ by noting that $\mathbf{A}_0$ is an orthogonal matrix. Thus
\begin{eqnarray}\label{C2}
\mathbf{MO}(\mathbf{T}(\mathbf{X}),F_{\mathbf{T}(\mathbf{X})})&=&[\lambda(\mathcal{I})]^{-1}\int_{\mathcal{I}} \mathbf{O}(\mathbf{T}(\mathbf{X}(t)),F_{\mathbf{T}(\mathbf{X}(t))}){\rm d}t\\ \nonumber
&=&[\lambda(\mathcal{I})]^{-1}\int_{\mathcal{I}} {f(t)\over \|f(t)\|}\mathbf{A}_0\cdot\mathbf{O}(\mathbf{X}(t),F_{\mathbf{X}(t)}){\rm d}t \nonumber \\
&=&\mathbf{A}_0\cdot\mathbf{MO}(\mathbf{X}(t),F_{\mathbf{X}(t)}). \nonumber
\end{eqnarray}
With (\ref{C1}) and (\ref{C2}), we get $\mathbf{MO}\left(\mathbf{T}(\mathbf{X}_g),F_{\mathbf{T}(\mathbf{X}_g)}\right)=\mathbf{A}_0\cdot\mathbf{MO}(\mathbf{X}(t),F_{\mathbf{X}(t)})$.

Consequently, we have ${\rm VO}(\mathbf{T}(\mathbf{X}_g),F_{\mathbf{T}(\mathbf{X}_g)})={\rm VO}(\mathbf{X}(t),F_{\mathbf{X}(t)})$.
Properties 2 and 3 are immediate results from the properties of $\|\mathbf{O}\|$ in Theorem \ref{Theorem 1}.\hfill $\Box$

\vskip 5pt
\noindent
\emph{Proof of Theorem \ref{Theorem 4}.}
\begin{eqnarray*}
	&&\sup_{\mathbf{X}\in (\mathbb{R}^p)^k}\|\mathbf{MO}_{T_k,n}\left(\mathbf{X},F_{\mathbf{X},n}\right)-\mathbf{MO}_{T_k}\left(\mathbf{X},F_{\mathbf{X}}\right)\|\\
	&=&\sup_{\mathbf{X}\in (\mathbb{R}^p)^k}\left\|{1\over k}\sum_{i=1}^k \left\{\mathbf{O}_n\left(\mathbf{X}(t_i),F_{\mathbf{X}(t_i),n}\right)w_n(t_i)-\mathbf{O}\left(\mathbf{X}(t_i),F_{\mathbf{X}(t_i)}\right)w(t_i)\right\}\right\|\\
	%&=&\sup_{\mathbf{X}\in (\mathbb{R}^p)^k}\left\|{1\over k}\sum_{i=1}^k \left\{\mathbf{O}_n\left(\mathbf{X}(t_i),F_{\mathbf{X}(t_i),n}\right)w_n(t_i)-\mathbf{O}\left(\mathbf{X}(t_i),F_{\mathbf{X}(t_i)}\right)w_n(t_i)\right\}\right.\\
	%&&\left.+{1\over k}\sum_{i=1}^k\left\{\mathbf{O}\left(\mathbf{X}(t_i),F_{\mathbf{X}(t_i)}\right)w_n(t_i)-\mathbf{O}\left(\mathbf{X}(t_i),F_{\mathbf{X}(t_i)}\right)w(t_i)\right\}\right\|\\
	&{\le} & \sup_{\mathbf{X}\in (\mathbb{R}^p)^k}\left\|{1\over k}\sum_{i=1}^k\left\{\mathbf{O}_n\left(\mathbf{X}(t_i),F_{\mathbf{X}(t_i),n}\right)-\mathbf{O}\left(\mathbf{X}(t_i),F_{\mathbf{X}(t_i)}\right)\right\}w_n(t_i)\right\|\\
	&&+\sup_{\mathbf{X}\in (\mathbb{R}^p)^k}\left\|{1\over k}\sum_{i=1}^k\mathbf{O}\left(\mathbf{X}(t_i),F_{\mathbf{X}(t_i)}\right)\left\{w_n(t_i)-w(t_i)\right\}\right\|.
\end{eqnarray*}
By convergence of $\mathbf{O}_n$ and $w_n(t_i)$, $t_i\in T_k$, we get the uniform, almost sure convergence of $\mathbf{MO}_{T_k,n}$ to $\mathbf{MO}_{T_k}$. \hfill $\Box$

\section*{Acknowledgments}
This research was supported by King Abdullah University of Science and Technology (KAUST). The authors
thank the editor, an associate editor, and three anonymous referees for their valuable comments.

\section*{References}

%\bibliography{directional_depth}

\begingroup
\setlength{\bibsep}{12pt}
\setstretch{1}
{
	\bibliography{directional_depth}
}
\endgroup

\end{document}